\documentclass[twocolumn]{aastex631} 
\usepackage{amsmath}
\usepackage[cm]{sfmath}

\newcommand{\mearth}{$\mathrm{M}_\oplus$}

\received{May 28, 2021}
\revised{January 26, 2022}
\accepted{March 17, 2022}

\submitjournal{PSJ}

\shorttitle{Impact ejecta and the late veneer}
\shortauthors{Carter \& Stewart}
\graphicspath{{./}{figures/}}

\begin{document}

\title{Did Earth eat its leftovers? Impact ejecta as a component of the late veneer}

\correspondingauthor{Philip J. Carter}
\email{p.carter@bristol.ac.uk}

\author[0000-0001-5065-4625]{Philip J. Carter}
\affiliation{School of Physics, H.H. Wills Physics Laboratory, University of Bristol, Bristol BS8 1TL, UK}
\affiliation{Department of Earth and Planetary Sciences, University of California, Davis, One Shields Avenue, Davis, CA 95616, USA}

\author[0000-0001-9606-1593]{Sarah T. Stewart}
\affiliation{Department of Earth and Planetary Sciences, University of California, Davis, One Shields Avenue, Davis, CA 95616, USA}


\begin{abstract}
The presence of highly siderophile elements in Earth’s mantle indicates that a small percentage of Earth’s mass was delivered after the last giant impact in a stage of `late accretion.' There is ongoing debate about the nature of late-accreted material and the sizes of {late-accreted} bodies. Earth appears isotopically most similar to enstatite chondrites and achondrites. It has been suggested that late accretion must have been dominated by enstatite-like bodies that originated in the inner disk, rather than ordinary or carbonaceous chondrites. 
Here, we examine the {provenance}s of `leftover' planetesimals present in the inner disk in the late stages of accretion simulations. Dynamically excited planet formation produces planets and embryos with similar {provenance}s, suggesting that the Moon-forming impactor may have had a stable isotope composition very similar to the proto-Earth. Commonly, some planetesimal-sized bodies with similar {provenance}s to the Earth-like planets are left at the end of the main stage of growth. The most chemically-similar planetesimals are typically fragments of proto-planets ejected millions of years earlier. If these similar-{provenance} bodies are later accreted by the planet, they will represent late-accreted mass that naturally matches Earth's composition. The planetesimal-sized bodies that exist during the giant impact phase can have large core mass fractions, with core {provenance}s similar to the proto-Earth. These bodies are an important potential source for highly siderophile elements. The range of core fractions in {leftover planetesimals} complicates simple inferences {as to the mass and origin of late accretion} based on the highly siderophile elements in the mantle.
\end{abstract}

\keywords{Earth -- impact phenomena -- planetesimals -- planet formation -- cosmochemistry -- planetary science -- meteorites -- solar system formation}


\section{Introduction}\label{s:intro}

The formation of the terrestrial planets in our solar system is usually considered to occur in a number of distinct (though likely overlapping) phases: planetesimal formation, runaway growth, oligarchic growth, and {the} culminating chaotic giant impact phase {\citep[e.g.][]{Liu20}}. It is generally accepted that the last major accretion event on the Earth -- the last giant impact -- was the event that led to the formation of the Moon {\citep[e.g.][]{Hartmann+Davis,Cameron+Ward,Cuk12,Canup12,Reufer12,Rufu17,Lock18}}. The Moon-forming impact is, however, not the {absolute} end to accretion; indeed, some accretion continues to this day. 

{The extant bodies of the solar system exhibit a wide range of isotopic compositions \citep[e.g.][]{Dauphas04,Fischer-Godde17}. In particular, there is extensive evidence across a range of elements for at least two substantially different isotopic reservoirs in the early solar system \citep[e.g.][]{Burkhardt11,Chen11,Warren11,Budde16,Scott18}: the `non-carbonaceous' reservoir typically associated with the inner solar system, and the `carbonaceous' reservoir typically associated with the outer solar system. None of the chondritic meteorites, often used as proxies for the building blocks of planets, are exact chemical and isotopic matches to Earth, although the enstatite chondrites have very similar isotopes. In some isotope systems the Earth is an end-member, and no mixture of other bodies can match the Earth's chemical and isotopic composition. The planets themselves show significant isotopic differences too, but, puzzlingly, the Moon is almost indistinguishable from the Earth in its isotopic composition \citep[e.g.][]{Wiechert01,Touboul07,Dauphas14}.}

Siderophile (`iron-loving') elements provide key insights into the properties of the growing Earth and the compositional characteristics of accreted bodies. Siderophile elements were variably removed from the mantle of the growing Earth by metal-silicate equilibration during episodes of core formation {\citep[e.g.][]{Wood06}}, and possibly via sulphide segregation {\citep[e.g.][]{ONeill91,Rubie16}}. The higher-than-expected abundances and chondritic proportions of highly siderophile elements (HSEs) in the Earth’s mantle suggest that a small percentage of the Earth's mass was delivered after the last giant impact in a stage of late accretion, also referred to as the `late veneer' \citep[e.g.][]{Walker09,Mann12}.

There is substantial debate about the {origin and bulk and isotopic composition} of this late-accreted material \citep[e.g.][]{Bermingham18,Hopp20}, and the sizes of the bodies that delivered it \citep[e.g.][]{Bottke10,Schlichting12,Brasser20}. The late veneer has commonly been {conflated with} delivery of volatile elements and compounds to the Earth, usually via `wet,' carbonaceous-chondrite-like or cometary bodies. It has, however, been {argued} that most of Earth's water arrived before the Moon-forming impact \citep{Halliday13,Dauphas13,Greenwood18}. Earth appears isotopically most similar to the enstatite chondrite family, and it has been suggested that late accretion must also have been dominated by enstatite-like material \citep{Dauphas17,Bermingham18}, rather than carbonaceous chondrites. The origin of late accreted bodies is unknown.  The lack of a substantial outer solar system chemical signature in the silicate Earth suggests that most late-accreted mass originated in the inner disk \citep[e.g.][]{Marty17}. Recent work by \citet{Fischer-Godde20} and \citet{Hopp20} suggests that late accretion consisted of a mixture of bodies, including carbonaceous-chondrite-like material and mass from an inner solar system reservoir not {recorded} in the meteorite record.

Since the HSEs measured in the Earth are in chondritic proportions, it has generally been assumed that {HSEs} were delivered by bodies of primordial {(undifferentiated)} composition. However, there is no requirement that these {bodies} were undifferentiated, chondrite-like {objects}. Chondrite parent bodies (or, at least, the unmelted outer layers of bodies from which chondrites may have originated) {formed} after the first generation of planetesimals, once planet growth {had} already begun \citep[e.g.][]{Dauphas11, Kruijer14}. If the late veneer consisted of differentiated planetesimals, or even small embryos \citep[e.g.][]{Raymond13,Genda17}, it is likely that at least some portion of the impactor's metallic cores would be delivered to the Earth's mantle, rather than sinking to the core {\citep[e.g.][]{Dahl10,Kendall16,Citron22}}. Thus, the late veneer need not consist of undifferentiated bodies, as long as their HSEs remain in approximately chondritic proportions overall.

One expected source of late veneer impactors is the population of planetesimals (or planetesimal-sized bodies) left in the inner solar system at the end of the main stages of accretion \citep[e.g.][]{Schlichting12}. In planet formation simulations, a number of planetesimal impacts occur at late times, after the final embryo-embryo giant impacts. In $N$-body simulations these planetesimals are typically assumed to be primordial, and their properties have not been examined closely. These {small} bodies may, however, have undergone significant collisional processing during the main stages of accretion, or represent fragments ejected from other collisionally evolved bodies \citep[e.g.][]{Genda17_GIFs}.

Here, we examine the {provenance}s of these `leftover’ planetesimals present in the inner disk in the late stages of accretion simulations. {Simulations of the intermediate stages of planet growth from previous works are analysed and detailed provenances of each body extracted. We find} that dynamically excited planet formation can result in terrestrial planets, embryos, and planetesimals with similar {provenance}s. We examine and quantify this similarity, and investigate the origins and evolution of these compositionally similar planetary bodies.


\section{Numerical methods}

The {planetary accretion} simulations discussed in this work were carried out using a modified version of the PKDGRAV $N$-body code \citep{Richardson00,Stadel01}. This code includes gravitational interactions between planetesimals (using a gravity tree), aerodynamic drag from the nebular gas, and planetesimal--planetesimal collisions. This version of PKDGRAV also includes the detailed empirically derived analytical collision model (EDACM) from \citet{Leinhardt12} and \citet{Leinhardt15}. EDACM provides realistic outcomes for collisions between planetesimals in the gravity-dominated regime. 

Most of the simulations discussed in this work were previously described in \citet{Carter15}; {one (022GTJf6hgas) is a new simulation that was conducted with the same version of the PKDGRAV code and the same standard parameters as used in \citeauthor{Carter15}}. We examine two dynamically contrasting scenarios: `calm accretion' -- with no giant planet perturbations, and the `Grand Tack' model -- in which Jupiter migrates inward through the protoplanetary disk and then back out toward its modern orbit. The migration began 2\,Myr after the start of the simulations. In both cases, simulations began with either 10\,000 or 100\,000 self-gravitating planetesimals distributed to produce a solid surface density comparable to the minimum mass solar nebula (MMSN) on `cold', near-circular orbits around a one solar mass star. These planetesimals have radii greater than 100\,km {(masses greater than $1 \times 10^{-5}$ \mearth)} and are therefore assumed to be differentiated into an iron core and silicate mantle from the start of the simulations.

{Our} simulations used radius enhancement to increase the collisional cross-section and speed up the evolution \citep[see][for further discussion]{Kokubo+Ida02,Carter15,Suli21}. 
{Radius enhancement reduces the time between collisions and thus changes the effective timescale of the simulations. While there are many bodies and frequent interactions radius enhancement does not affect the accuracy of the simulations \citep{Kokubo+Ida96}. Collisions close to escape velocity will have reduced impact velocities due to the enhanced radii \citep{Carter15,Suli21}, which may lead to a small underestimate of the degree of collisional erosion. } 
With particle radii enhanced by a factor of 6, the 600\,000\,year integration time represents an effective duration of $\sim$21\,Myr. Thus, the simulations evolve the protoplanetary disk through the runaway and oligarchic growth phases, but end before the conclusion of the giant impact stage of terrestrial planet formation. 

{PKDGRAV uses a second-order integrator; this integrator is fast, but lacks the accuracy of the higher-order methods typically used for simulations of the giant impact stage \citep[e.g.][]{Duncan98,Chambers01}. While the second-order integrator is advantageous during the early and intermediate stages of growth, when there are many bodies and many interactions, it is unclear whether the accuracy would be sufficient to model the long-term orbital evolution of the smaller numbers of bodies present during the bulk of the giant impact phase.} 

The properties of planetesimals and other planetary bodies at the end of these simulations could be considered representative of the {initial} compositions of bodies in {`classical'} simulations of the giant impact phase of planet formation \citep[e.g.][]{Chambers01,Walsh11}, {though care must be taken due to differing evolutionary timescales across the disk \citep{Carter15,Walsh19}}. The effective times should not be considered to be precise; however, this imprecision is not important for the outcomes of collisions or the compositional evolution that occurs during accretion.

The collision model, EDACM, calculates the size and velocity distribution of post-collision remnants and fragments {(third largest and smaller post-collision bodies)} from the standard collision parameters: impact angle, mass ratio, and impact velocity \citep[see][]{Leinhardt12,Stewart12,Leinhardt15}. When using a collision model that can produce fragments in an $N$-body code it is necessary to prevent production of increasing numbers of small fragment particles, usually by imposing a mass limit on fragments. Several authors have placed any remaining fragment mass back onto the largest post-collision remnant, or distributed this mass between a small number of fragments \citep[e.g.][]{Chambers13}. In PKDGRAV, mass assigned to fragments that would fall below this mass (or resolution) limit, $M_\mathrm{min}$, is instead placed in a debris annulus corresponding to the location in the disk at which the collision occurred. This `unresolved debris' is distributed in 0.1\,au wide annular bins throughout the modeled region of the disk. The unresolved debris consists of both `core' and `mantle' components. As well as being generated in collisions, the debris is re-accreted by resolved planetesimals on non-circular orbits passing through each annulus \citep{Leinhardt05,Carter15,Leinhardt15}. The small scale debris is thus recycled and the instantaneous mass in debris never grows to a large fraction of the total mass, even though a large fraction of the total mass is processed through this debris \citep{Bonsor15,Carter15,Davies20}.

After each collision, the core and mantle components of the impactors are distributed to post-collision remnants {and fragments} according to a mantle stripping law based on \citet{Marcus10}. The simulations discussed in this work used mantle stripping `model 3' as defined in \citet{Carter15} as the average of the two models given by \citeauthor{Marcus10}. We note that the way core material was distributed to smaller post-collision {bodies} in these simulations -- assigning core mass first to `fill' remnants{/fragments} in descending mass order -- may overproduce high core mass fraction bodies \citep[see][]{Carter18}. Core and mantle masses remain distinct throughout the simulations; we neglect chemical exchange and re-equilibration.

The migration of Jupiter in the Grand Tack model is achieved by imposing an acceleration on the particle representing the giant planet. The planetesimals experience aerodynamic drag from the nebular gas based on the prescription of \citet{Adachi76}; 
{ calculated using natural, not enhanced, radii. } 
The nebular gas surface density profile is based either on the MMSN, or a denser gas disk with gaps produced by the giant planets based on \citet{Morbi07} as used in several previous works \citep[e.g.][]{Walsh11,Raymond17,Carter15,CarterStewart20} -- which we refer to in the rest of this work as the MC gas profile.   
After $\sim$2\,Myr (the time corresponding to giant planet migration), the nebula dispersed with an e-folding timescale, $\tau_\mathrm{gas}$, of 0.1\,Myr in most simulations. {In two of the calm accretion simulations the nebular gas surface density remained unchanged throughout the simulation run.}

A summary of the simulations discussed in this work is given in Table \ref{t:sims}. Further details on all aspects of the simulations can be found in \citet{Carter15}.

\begin{table*}
    \centering
    \caption{Summary of PKDGRAV planet formation simulations examined in this work. \label{t:sims}}
    \begin{tabular}{lcrcccccc}
    \hline
    \hline
    Simulation & Type & $N$ & Outer edge (au) & Disk mass (M$_\oplus$) & Core fraction & Gas disk & $\tau_\mathrm{gas}$ (kyr) & $M_\mathrm{min}$ (M$_\oplus$)\\
    \hline
    022GTJf6hgas & GT & 100\,000 & 3.0 & 4.85 & 0.22 & MC & 100 & 2$\times 10^{-5}$ \\
    022GTJf6hgas\_2 (27) & GT & 10\,000 & 3.0 & 4.85 & 0.22 & MC & 100 & 2$\times 10^{-4}$ \\
    035GTJf6hgas (28) & GT & 10\,000 & 3.0 & 4.85 & 0.35 & MC & 100 & 2$\times 10^{-4}$ \\
    022f6sgas\_3 (14) & calm & 100\,000 & 1.5 & 2.5 & 0.22 & MMSN & -- & 1$\times 10^{-5}$ \\
    022f6sgas\_4 (18) & calm & 100\,000 & 1.5 & 2.5 & 0.22 & MMSN & 100 & 1$\times 10^{-5}$ \\
    035f6sgas (15) & calm & 100\,000 & 1.5 & 2.5 & 0.35 & MMSN & -- & 1$\times 10^{-5}$ \\
    035f6sgas\_2 (19) & calm & 100\,000 & 1.5 & 2.5 & 0.35 & MMSN & 100 & 1$\times 10^{-5}$ \\
    \hline
    \end{tabular}\medskip\\
    {Note: The numbers in parentheses after the simulation name are the simulation IDs given in table 4 of \citet{Carter15}; $N$ is the initial number of planetesimals; the inner edge of the planetesimal disk was 0.5\,au in all cases; $\tau_\mathrm{gas}$ is the gas dissipation timescale and giant planet migration timescale, `--' indicates the gas density remained constant throughout; $M_\mathrm{min}$ is the mass resolution limit; {and M$_\oplus$ is the mass of the Earth.}}
\end{table*}

{\subsection{Tracking provenance}}

\begin{figure*}
\centering\includegraphics[width=\textwidth]{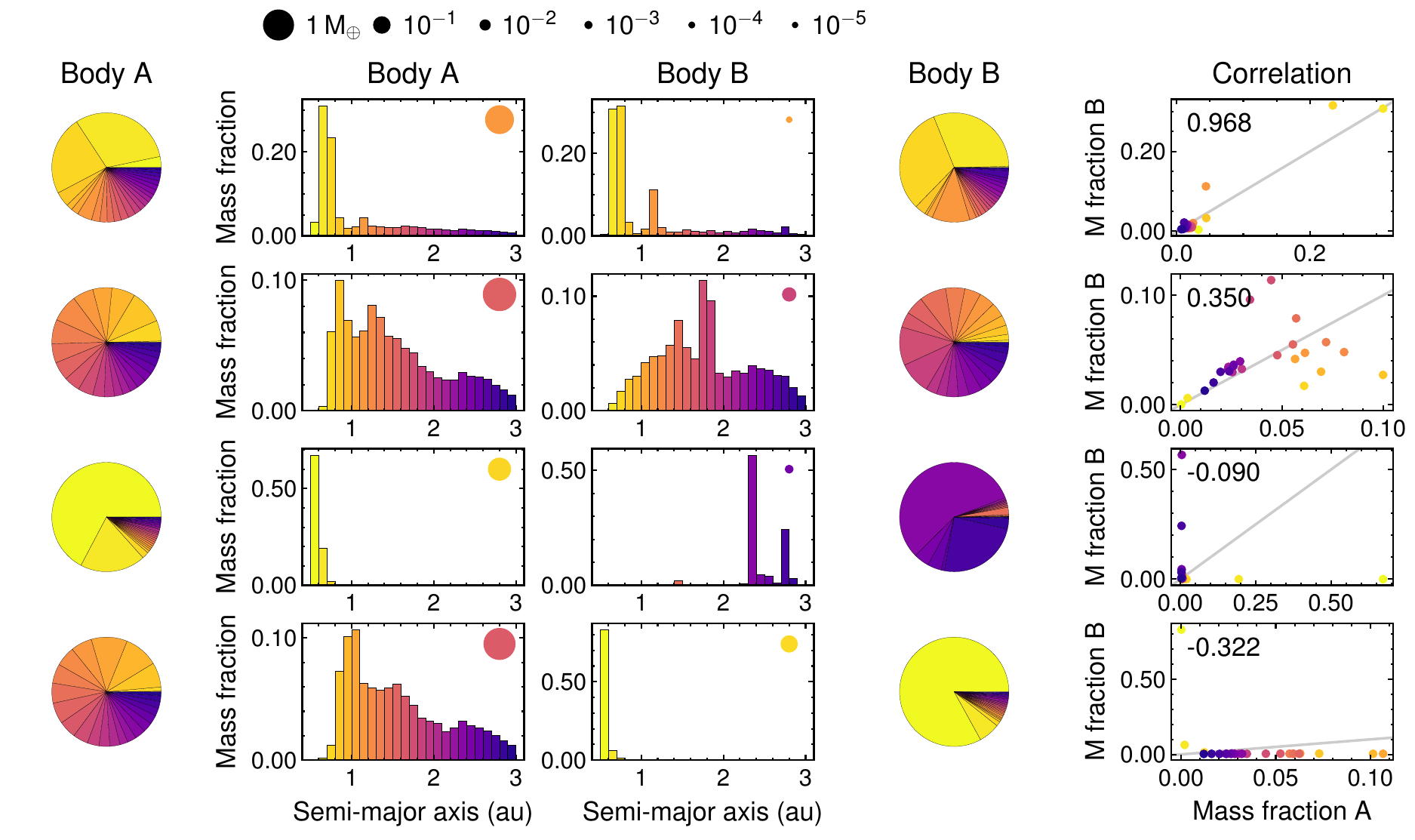}
\caption{Examples of correlations between pairs of planetary bodies formed in a Grand Tack simulation (022GTJf6hgas). The leftmost column shows the {provenance} of an embryo (body A) as a pie chart, the second column shows the corresponding histogram as a function of heliocentric distance. The third and fourth columns show the histogram and pie chart {provenance} for the comparison body (B). The final, rightmost column indicates the correlation of the bodies' {provenance}s with the mass fraction of the two bodies originating in each radial bin plotted against each other. The grey line is the 1:1, A=B line, and is shown for reference. The number in the top left corner of the rightmost panels is the calculated Pearson correlation coefficient for the pair of bodies. The colored circles in the upper right corner of the histogram panels indicate the relative sizes of the bodies {(according to the scale at the top of the figure)}, and are colored according to the average color of the body's {provenance} histogram. \label{f:ccexample}}
\end{figure*}

{The provenance of a body indicates the radial location in the disk from which its mass originated. Provenance histograms correspond to the same 0.1\,au wide annuli as the unresolved debris bins.} 
All bodies begin with {their provenance (numerical composition) defined according to their} initial radial location.  
This tracking scheme is motivated by the idea of radial gradients {or zoning} in the initial bulk and isotopic composition of solids in the disk {\citep[e.g.][]{Dauphas04,Fischer-Godde17}}. 
{The provenance of a body evolves} any time {the} body experiences a collision or accretes debris. Unlike in most other $N$-body simulations, in this work, planetesimals can accrete both via collisions with other resolved bodies and via unresolved debris. Remnants and fragments of collisions are updated to have the mass-weighted average {provenance} of the impactors (examples of these {provenances} are shown in Figure \ref{f:ccexample}). {Each unresolved debris bin also has its provenance histogram averaged with newly produced debris (weighted by mass) after each collision that occurs in that annulus.}
This model does not include the formation of new planetesimals directly from dust or pebbles. 

We extracted the {provenance} histograms -- arrays of mass fraction derived from each radial bin -- for all bodies at regular intervals throughout the simulations. The {provenance}s of the metallic cores of these differentiated bodies were also extracted. These {provenance}s can be compared by calculating the Pearson correlation coefficient of the {provenance} histograms of any pair of bodies. Examples of {provenance}s and correlations are given in Figure \ref{f:ccexample}. A correlation coefficient of $+1$ indicates a perfect match in {provenance}; the first row of Figure \ref{f:ccexample} shows a very close match with a coefficient of $+0.97$. Large negative values occur when most of the radial bins have substantially differing contributions between the two bodies. For example, the final row in Figure \ref{f:ccexample} shows one embryo with substantial material from all but the first bin, and a second embryo that acquired the majority of its mass from the first bin, thus leading to a correlation coefficient of $-0.32$. Perfect anti-correlation {($-1$)} is very unlikely to occur. {As well as the full mass-origin distribution or `provenance histogram' for a body, we can also calculate a mass-weighted mean origin distance for each body, which we represent by the color corresponding to that original location in the disk.}  The top left embryo in Figure \ref{f:ccexample} acquired most of its mass from yellow-colored bins interior to 0.8\,au, but its average color is orange -- corresponding to $\sim$1.1\,au -- due to the contribution of material from greater heliocentric distances.

Since we track each body's mass and {provenance} throughout the simulation, we can thus determine how similar any planetesimal is to each embryo as a function of time. In this work we define embryos as bodies with masses of at least 1\% the mass of the Earth, and planetesimals as those bodies smaller than this limit. There is no computational distinction between these size classes.


\section{Results}

\begin{figure*}
\centering\includegraphics[width=\textwidth]{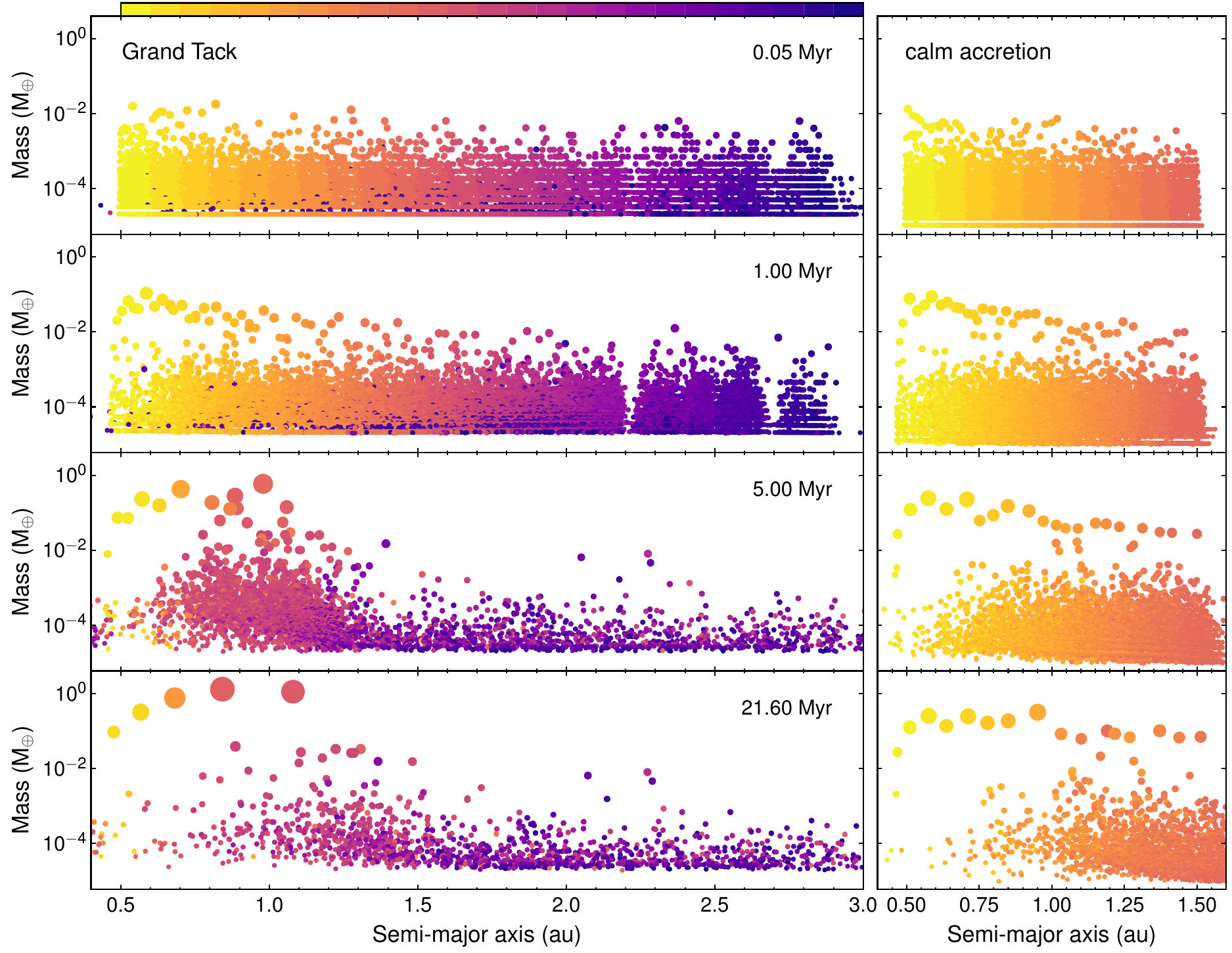}
\caption{Mass distribution as a function of semi-major axis and time for bodies in a Grand Tack simulation (left, 022GTJf6hgas) and a calm accretion simulation (right, 022f6sgas\_4). The sizes of the points are scaled to the mass of each body. Colors indicate the mean origin distance of mass making up the body, with the corresponding initial colors shown in the bar above the top left panel. {The time shown in each row is indicated in the upper right of the left hand panel.} The evolution of the inner disk is very similar for the first two time {rows}, but after the tack (at 2\,Myr), the mass and {provenance} (color) distribution in the Grand Tack scenario has substantially diverged from that in the calm accretion model.\label{f:massdist}}
\end{figure*}

We begin with a brief overview of the general properties of the proto-planetary systems formed in these simulations before we delve deeper into their {provenances}. For a more detailed discussion of the outcomes of the simulations see \citet{Carter15}.

\subsection{Proto-planetary disk evolution}

{A disk of planetesimals is expected to undergo a phase of runaway growth in which the most massive planetesimals grow in mass quicker than smaller planetesimals. The few largest planetesimals thus detach from the continuous power law size distribution of the smaller bodies \citep[e.g.][]{Kokubo+Ida96}. As the larger planetesimals continue to grow into planetary embryos they begin to significantly excite the orbits of the surrounding population of small planetesimals, halting runaway growth. The continued growth of embryos remains faster than the growth of small planetesimals, but the largest embryo grows more slowly than smaller embryos. The masses of the planetary embryos thus become closer as growth proceeds -- this is known as oligarchic growth \citep{Kokubo+Ida98}. In our simulations,} 
runaway planetesimal growth gives way to oligarchic growth of planet embryos in an `inside-out' progression (Figure \ref{f:massdist}). The disk surface density is higher and orbital timescales are shorter closer to the star; thus evolution is faster closer to the star. At 20\,Myr, in the calm accretion scenario the majority of planetesimals have been accreted at 0.6\,au, and Mars-sized embryos dominate; while there are still large numbers of planetesimals at 1.5\,au, and Mars-sized embryos have only begun to form {\citep[see also][]{Walsh19}}.

The Grand Tack leads to some substantial changes in this evolution. As expected, Jupiter's migration truncates the inner disk, emplacing a substantial mass of material that was in its path into the terrestrial planet region. The early increase in surface density accelerates the evolution. While inside-out growth is still evident, embryos grow more massive and more rapidly than in the calm scenarios (see Figure \ref{f:massdist}). Planetesimal and embryo eccentricities and impact velocities are higher across the inner solar system in the dynamically hot Grand Tack scenarios. The higher eccentricities and impact velocities cause greater mantle stripping, and lead to greater diversity in core-mantle ratio of planetary embryos in the Grand Tack scenarios \citep{Carter15}. No mass is lost completely in these simulations, and both resolved and unresolved material can be accreted; thus mantle stripping leads to diversity in core fraction rather than global increase.

In the Grand Tack, Jupiter shepherds a large fraction of planetesimals from the 2--3\,au region into the region of the disk where the Earth is growing. This shepherding substantially increases the mixing of material that originated at different distances from the Sun, and leads to substantially different embryo {provenance}s than the calm accretion model (see also \citealt{Carter15} fig. 14).

\subsection{Provenance}

\begin{figure*}
\centering\includegraphics[width=0.82\textwidth]{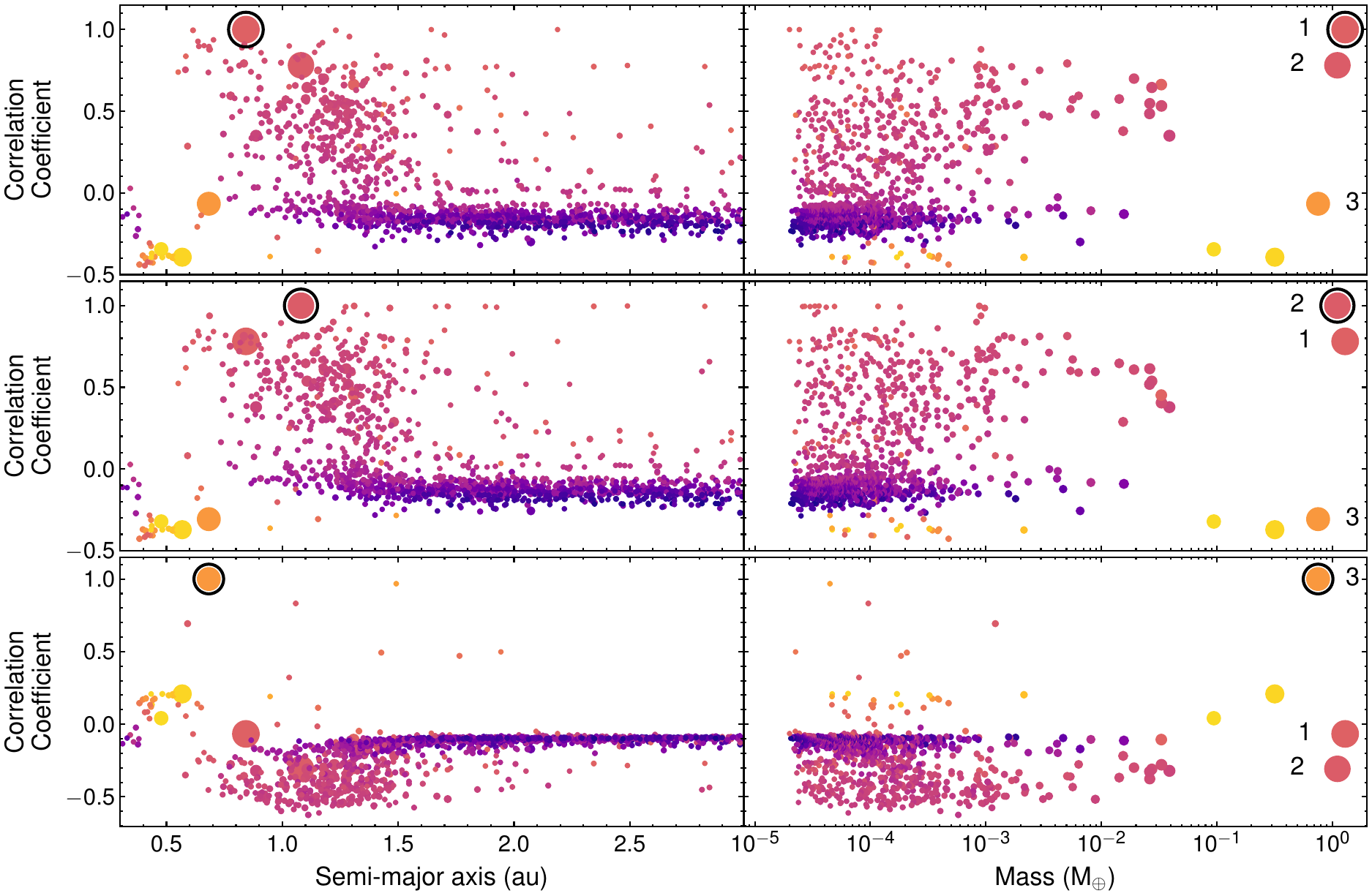}
\caption{Correlation of bodies to the three most massive embryos (highlighted {with black outlines} and labelled in each row) {as a function of semi-major axis and body mass} at the end of an example Grand Tack simulation (022GTJf6hgas). The sizes of the points are scaled to the mass of each body. Colors indicate the mean origin distance of mass making up the body as in Figure \ref{f:massdist}. \label{f:embcompareGT}}
\end{figure*}
\begin{figure*}
\centering\includegraphics[width=0.82\textwidth]{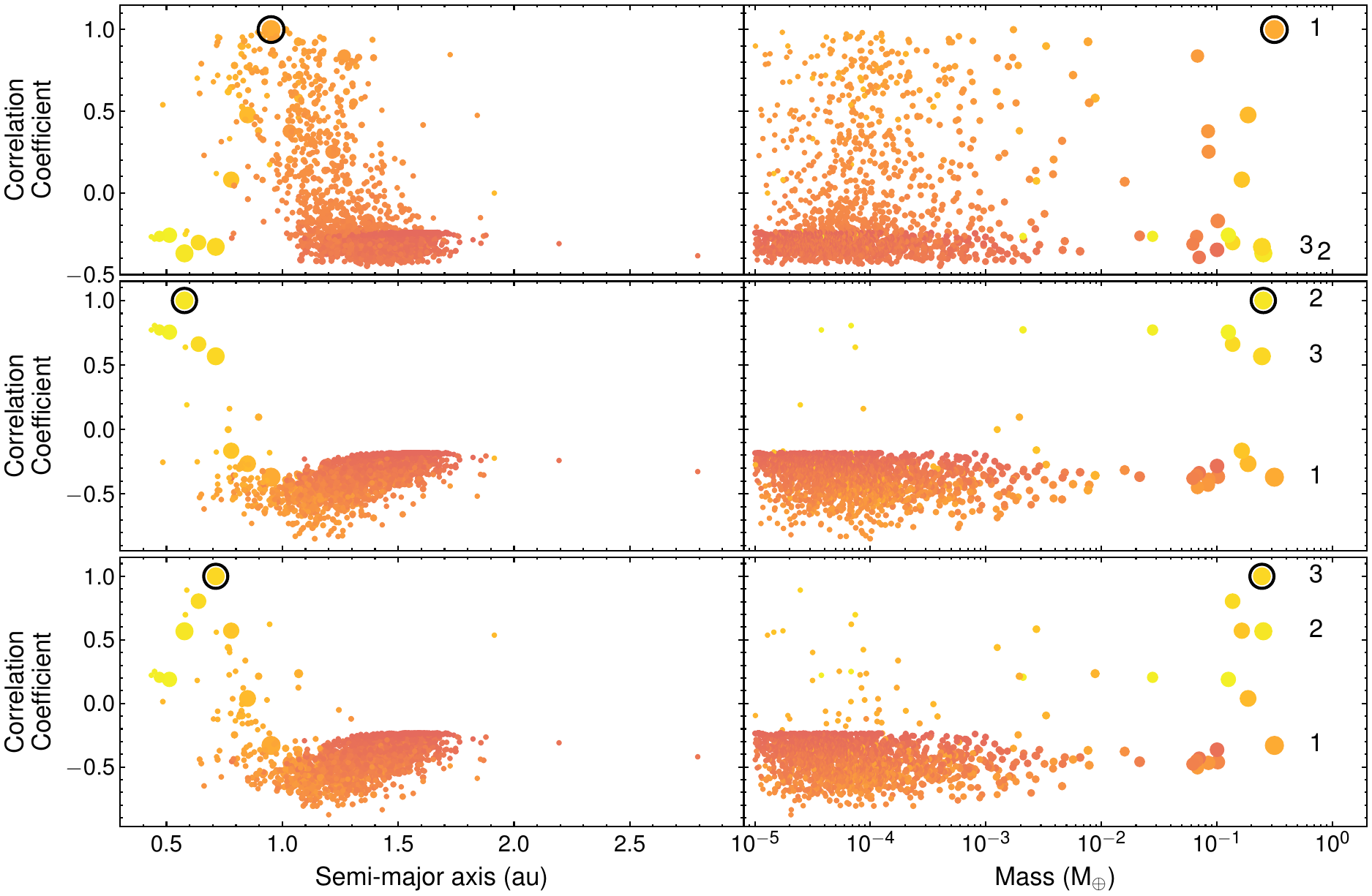}
\caption{Correlation of bodies to the three most massive embryos (highlighted {with black outlines} and labelled in each row) {as a function of semi-major axis and body mass} at the end of an example calm simulation (022f6sgas\_4). The sizes of the points are scaled to the mass of each body. Colors indicate the mean origin distance of mass making up the body as in Figure \ref{f:massdist}. \label{f:embcomparecalm}}
\end{figure*}

In this work, {provenance} {refers} to the distribution in original radial location of the materials accreted by a body. The simulations discussed here allow a more detailed comparison than the classical embryo `feeding zone' analysis {\citep[e.g.][]{Raymond09,Kaib15}}. {There is abundant evidence for at least two distinct isotopic reservoirs in the early solar system \citep[e.g.][]{Burkhardt11,Chen11,Scott18}, and the protoplanetary disk may have possessed a continuously varying distribution of elements or isotopes with radial distance \citep[e.g.][]{Lewis74,Dauphas04,Qin09,Trinquier09,Fischer-Godde17}. } We do not assign specific chemical or isotopic signatures to the radial bins in our simulations, nor a chemical gradient across the disk. Naturally, chemical differences between regions of the disk is what we allude to, but without detailed knowledge of the original distribution we cannot meaningfully assign specific, detailed properties to radial bins. In the most general terms, bluer (purple and blue) material originates farther from the star and is likely, therefore, to be more volatile-rich than material from yellow and orange bins; but we impose no fixed model on this radial composition. 

{The average colors (mean origin distance) seen} in Figures \ref{f:massdist}, \ref{f:embcompareGT}, and \ref{f:embcomparecalm} {show} a clear difference in {provenance}s and compositional similarities between the two example simulations. At the end of calm accretion simulations embryos generally have average colors ({provenance}s) that correspond closely to their locations. In Grand Tack simulations most of the embryos (and planetesimals) are substantially `bluer' due to the contribution of material shepherded inward by Jupiter's migration.

The dynamically excited Grand Tack simulations produce Earth-like proto-planets with a mixture of material from across the inner solar system (see Figure \ref{f:ccexample}). This {mixing} is a result of Jupiter's inward migration `pushing' planetesimals from more distant parts of the inner disk into the Earth-forming region. As can be seen in Figures \ref{f:massdist} and \ref{f:embcompareGT}, there are multiple smaller embryos remaining at the end of these simulations, several of which have fairly similar {provenance}s to the larger bodies. The calm accretion model produces embryos that are dominated by material originating close to their locations, resulting in a lower degree of similarity between the final embryos (Figure \ref{f:embcomparecalm}).

\begin{figure*}
\centering\includegraphics[width=\textwidth]{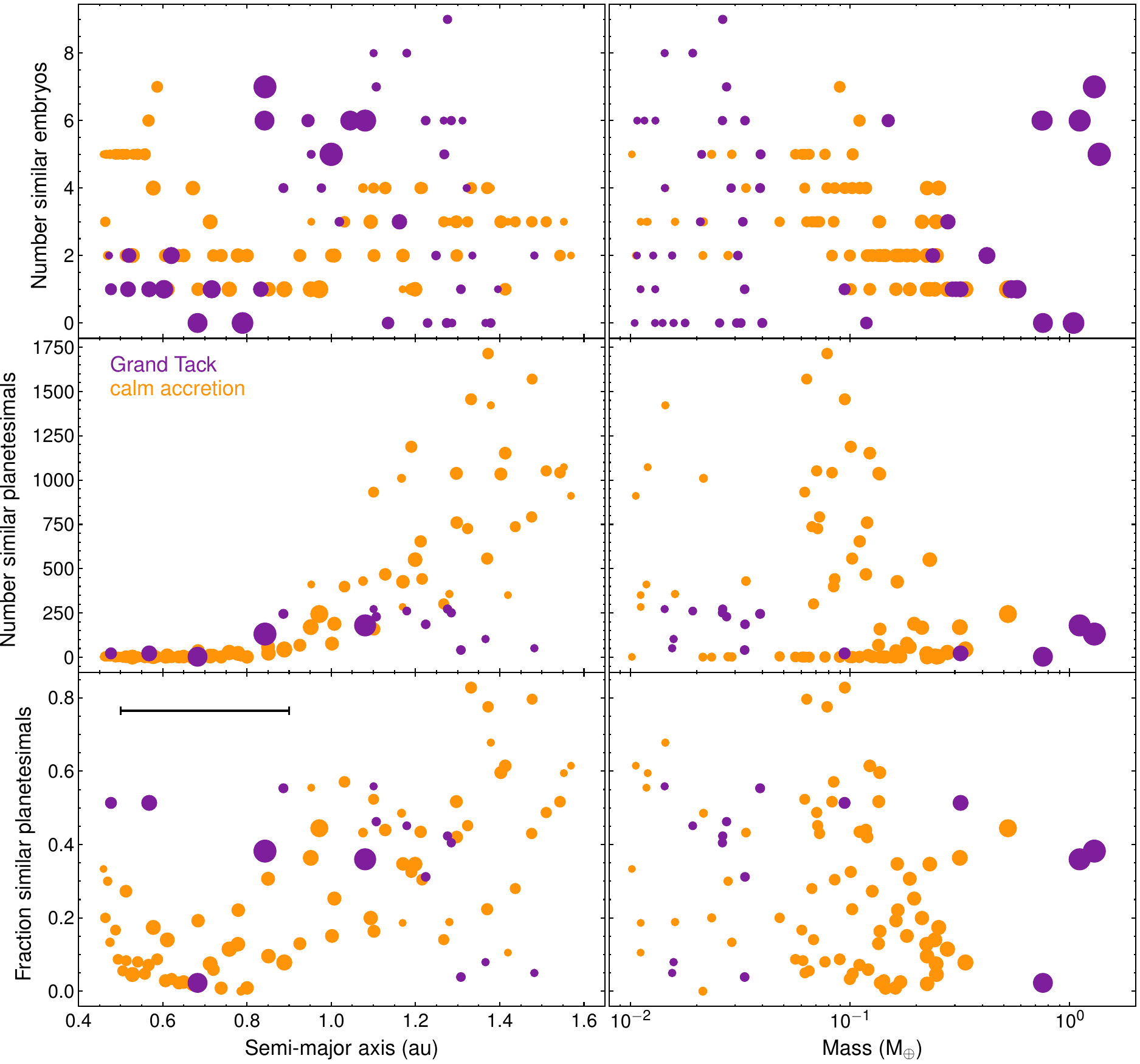}
\caption{Number of embryos (M$\geq$0.01\,\mearth, {top}), {number of} planetesimals (M$<$0.01\,\mearth, {middle}), {and fraction of nearby planetesimals (bottom)} with correlation coefficients of 0.5 or above for each embryo in several simulations. The numbers of similar bodies are shown as a function of semi-major axis (left) and mass {of the relevant embryo} (right). {Purple} points are for Grand Tack simulations, {orange} points are for calm accretion simulations. {The fraction of planetesimals that are similar to each embryo is calculated only for planetesimals with semi-major axes within 0.2\,au of the embryo, as indicated by the black horizontal bar in the lower left panel.} The sizes of the points are scaled to the mass of the embryos. Embryos from low resolution simulations are excluded from the planetesimal (lower) panels. \label{f:nsimilar}}
\end{figure*}
Figures \ref{f:embcompareGT} and \ref{f:embcomparecalm} reveal a wide range of planetesimal {provenance}s at the end of these simulations. Embryos in the innermost regions of the disk have fewer planetesimals with similar {provenance}s compared to embryos that formed closer to 1\,au {due to the faster evolutionary timescales closer to the star}. 

{\subsection{Provenance similarity}}

\begin{figure*}
\centering\includegraphics[width=\textwidth]{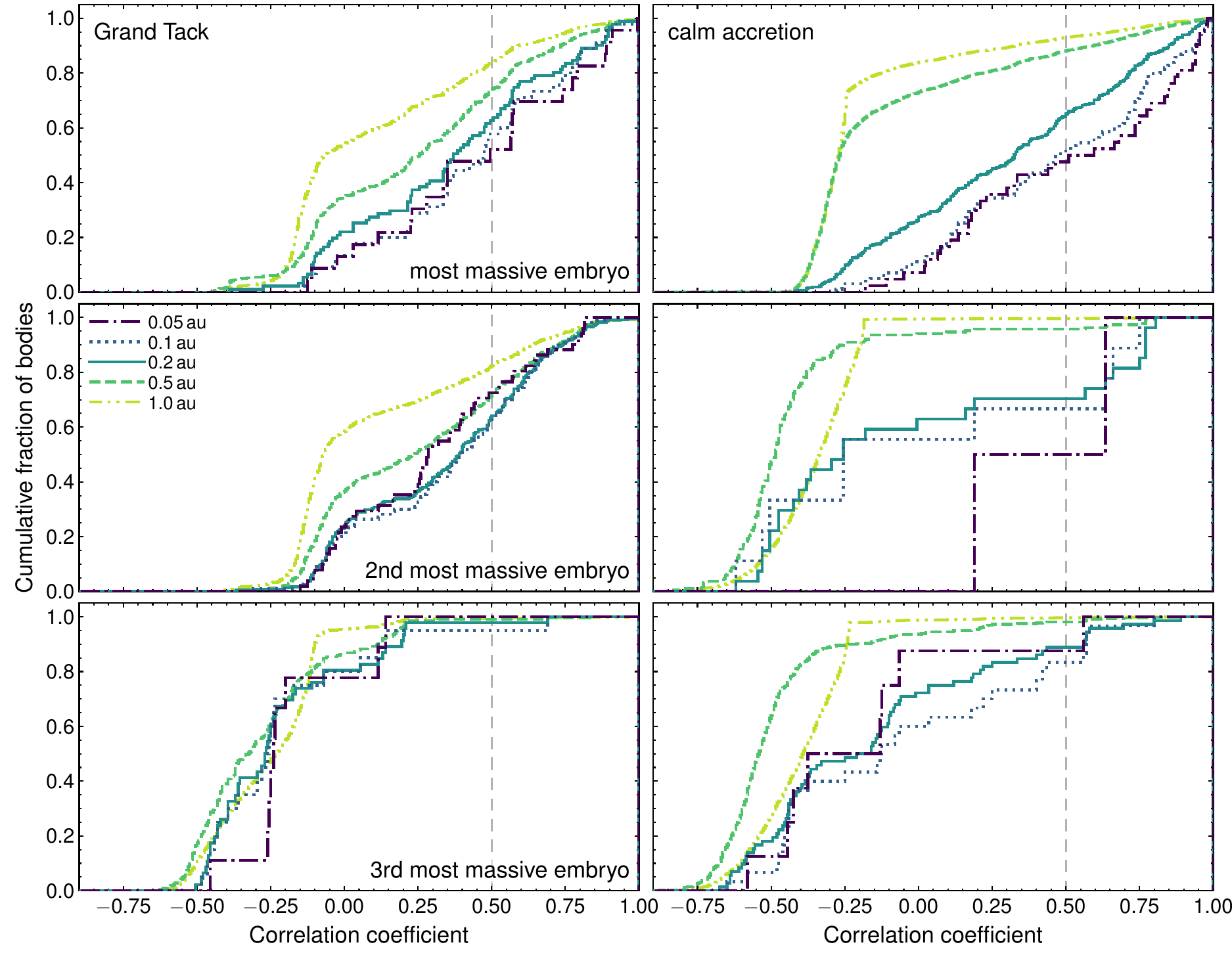}
\caption{{Provenance correlation distributions for nearby bodies with increasing distance from the embryo for the three most massive embryos in example Grand Tack (022GTJf6hgas) and calm accretion (022f6sgas\_4) simulations. The colors and line styles (see legend) indicate the width of the bin around the embryo in which correlation is calculated. 
The grey dashed vertical line highlights the correlation coefficient of 0.5 used for comparison elsewhere in this work.} \label{f:hists}}
\end{figure*}
The numbers of similar bodies to each final embryo from all the simulations are shown in Figure \ref{f:nsimilar}. Each point in Figure \ref{f:nsimilar} represents an individual embryo, with the $y$-axis values showing the number of other embryos or planetesimals similar to it at the end of the simulation in which it formed. The top row shows the number of other embryos with a {provenance} correlation coefficient of at least 0.5 for each embryo. The {middle} row shows the numbers of planetesimals with a {provenance} correlation coefficient of at least 0.5 for each embryo in the high resolution simulations. {The bottom row shows the fraction of planetesimals with final semi-major axes within 0.2\,au of the embryo that have a {provenance} correlation coefficient of at least 0.5 for each embryo in the high resolution simulations.} The choice of 0.5 is somewhat arbitrary; the distribution is similar with correlation cutoff between 0.2 and 0.7. The left-hand panels show embryos as a function of their final semi-major axis, the right-hand panels show the embryos as a function of their final mass. {Purple} points show embryos formed in the set of Grand Tack simulations and {orange} points show embryos from the calm accretion simulations. 

We see that there are typically greater numbers of similar embryos (5--8) for the larger embryos produced near 1\,au in the Grand Tack simulations. The most massive embryos in calm accretion simulations have already accreted much of the material with similar {provenance}s, and so there are typically fewer similar embryos remaining at, and interior to, 1\,au. 

At the end of these Grand Tack simulations, there are several smaller embryos in the region close to 1\,au that are very similar in {provenance} to the Earth-like proto-planets (Figures \ref{f:embcompareGT} and \ref{f:nsimilar}). It is expected that these embryos will be removed during the subsequent $\sim$100\,Myr via collision or ejection. One of these smaller embryos may represent Theia -- the body that collided with the proto-Earth leading to the formation of the Moon. The similar {provenance}s of the proto-Earth and these smaller embryos suggest that Theia may have had a very similar stable isotope composition to the proto-Earth. Similar compositions for Theia and the proto-Earth would likely have resulted in an Earth and Moon with near-identical stable isotope compositions.

The numbers {and fractions} of planetesimals similar to embryos shown in Figure \ref{f:nsimilar} have noticeably distinct distributions compared to the numbers of similar embryos {(top row)}, with a very clear difference between the two evolution scenarios. For calm accretion the number of similar planetesimals to each embryo shows a clear and straightforward trend of increasing numbers of similar planetesimals with increasing semi-major axis. This {pattern} reflects the inside-out evolution of the planetesimal disk. Closer to the star, where orbital timescales are shorter, embryos grow faster and accrete the majority of planetesimals from their feeding zone. At greater distances from the star accretion is slower, there are more planetesimals remaining, {and embryos have not grown as massive}. Without substantial mixing of the disk nearby planetesimals are very likely to be similar to embryos growing in the same region.

In the Grand Tack scenarios, inside of $\sim$0.8\,au {many} planetesimals have already been accreted by $\sim$5\,Myr, resulting in few planetesimals similar to embryos in the innermost regions of the disk (Figure \ref{f:nsimilar}). Jupiter's tack substantially mixes the disk, excites orbits, and concentrates planetesimal mass in an annulus between $\sim$0.8 and 1.3\,au. These effects of the Grand Tack, combined with collisional evolution, result in many planetesimals with similar {provenance}s to the embryos remaining in this annulus region at 20\,Myr {(130--270 planetesimals; $\sim$40\% of those within 0.2\,au)}. The more rapid growth of embryos in the inner-most region of the disk, and the lower degree of mixing from Jupiter's migration, leave the `Mercury' region very different to the Earth region of the disk. This inner-most part of the disk has fewer similar bodies, and {provenance}s closer to the initial {provenance}s of the annuli, similar to the calm accretion scenario. At larger distances from the star there is less mass left in the disk (fewer planetesimals and fewer embryos), leading to the decrease in numbers of planetesimals similar to embryos beyond $\sim$1.3\,au seen in Figure \ref{f:nsimilar}.

{The bottom panel of Figure \ref{f:nsimilar} shows the fraction of similar planetesimals within 0.2\,au of each embryo. We see distributions in semi-major axis similar to those for the number of similar planetesimals but with greater variation. Figure \ref{f:hists} shows the fraction of bodies as a function of correlation coefficient and separation for the three most massive embryos from two example simulations. The solid cyan line corresponds to a region extending 0.2\,au either side of the embryo as used in Figure \ref{f:nsimilar}. The Grand Tack scenario shows little variation in similarity with distance from the embryo, while there is a much larger decrease in similarity in the calm accretion case. In the Grand Tack example (022GTJf6hgas) $\sim$20\% of bodies within 1\,au have a provenance correlation of 0.5 or greater for the two most massive embryos; however only 7\% and 0.5\% of bodies within 1\,au are as similar for the two most massive embryos from the calm accretion example (022f6sgas\_4). }

\subsection{Origins of similar planetesimals}

We have seen that planetesimals with similar {provenance}s to planetary embryos can be present in the disk early in the giant impact phase of evolution. We now examine the histories of these small bodies to learn how and why they {became} similar in {provenance} to the embryos.

\begin{figure*}
\centering\includegraphics[width=\columnwidth]{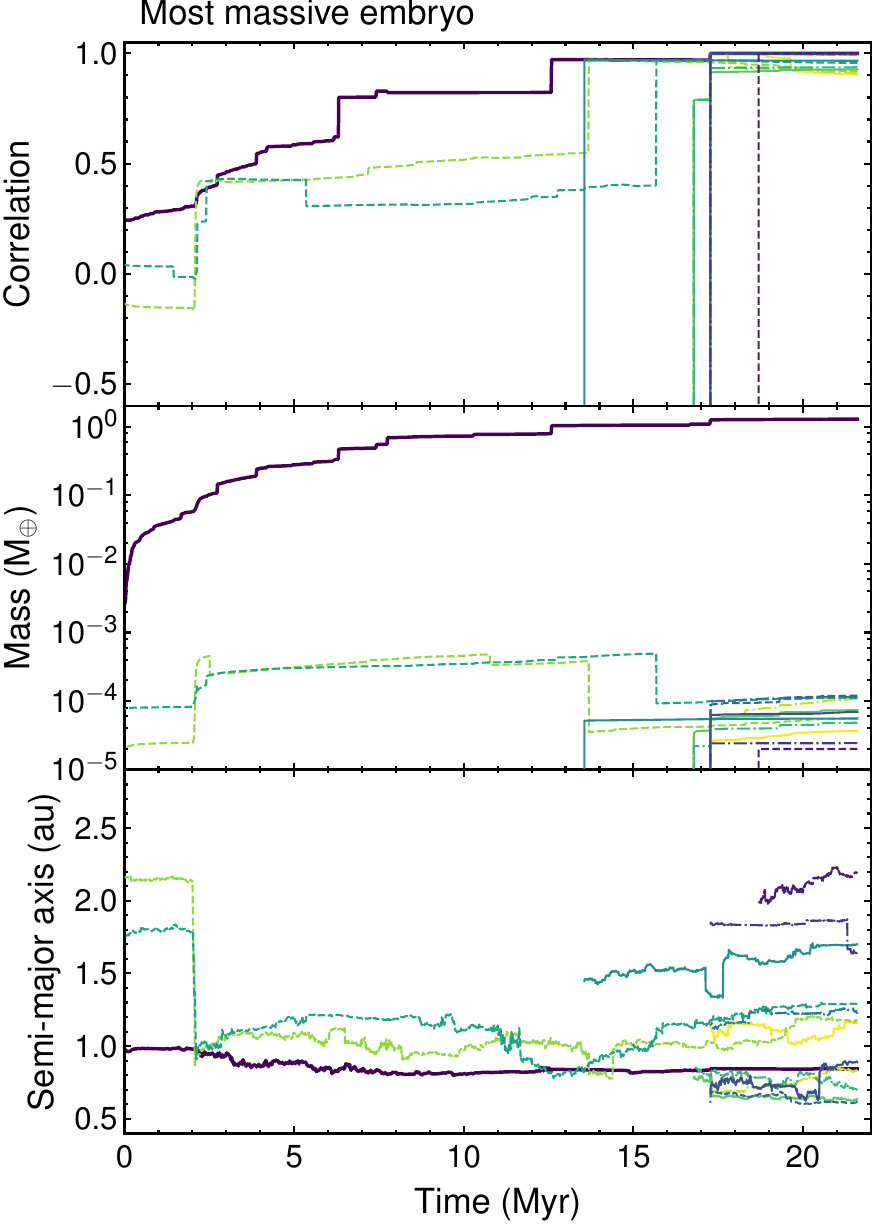}\hspace{8mm}
\centering\includegraphics[width=\columnwidth]{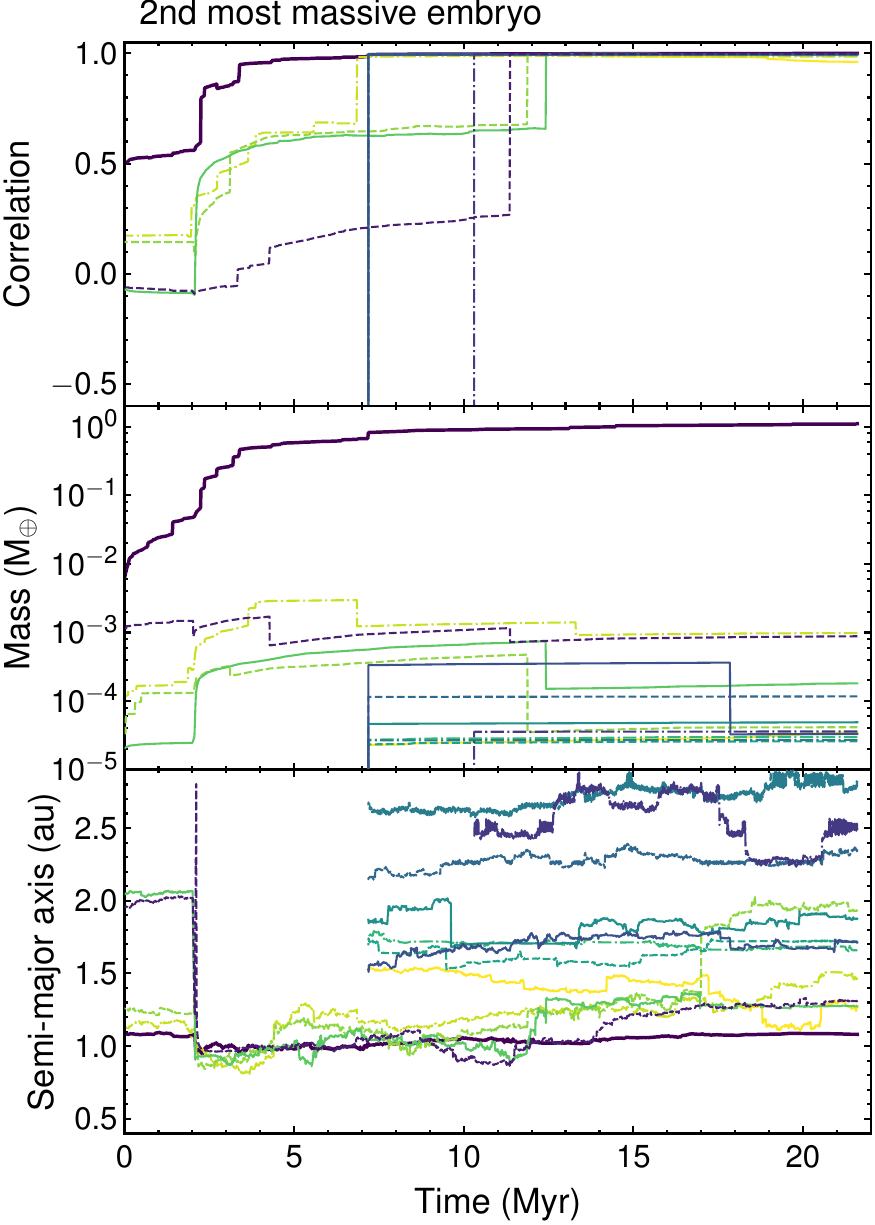}
\caption{Evolution of {provenance} correlation coefficient,  body mass, {and semi-major axis} for the growing embryo and the 12 bodies with the most similar {provenance}s at the end of an example Grand Tack simulation (022GTJf6hgas). The {left} and {right} plots correspond to the most massive and second most massive embryos as shown in Figure \ref{f:embcompareGT}. The thick dark purple line indicates the embryo and the thin solid, dashed, and dot-dash lines the most similar bodies, colored in order of correlation coefficient (purple -- highest -- to yellow -- lowest). 
{Note that the properties of bodies shown in these plots, including orbits, correspond to the fixed output times of the provenance data; collision ejecta may have undergone tens of orbits before appearing on the plots, hence the wide range of semi-major axes for new fragments.} 
\label{f:correlationevolutionGT}}
\end{figure*}

\begin{figure*}
\centering\includegraphics[width=\columnwidth]{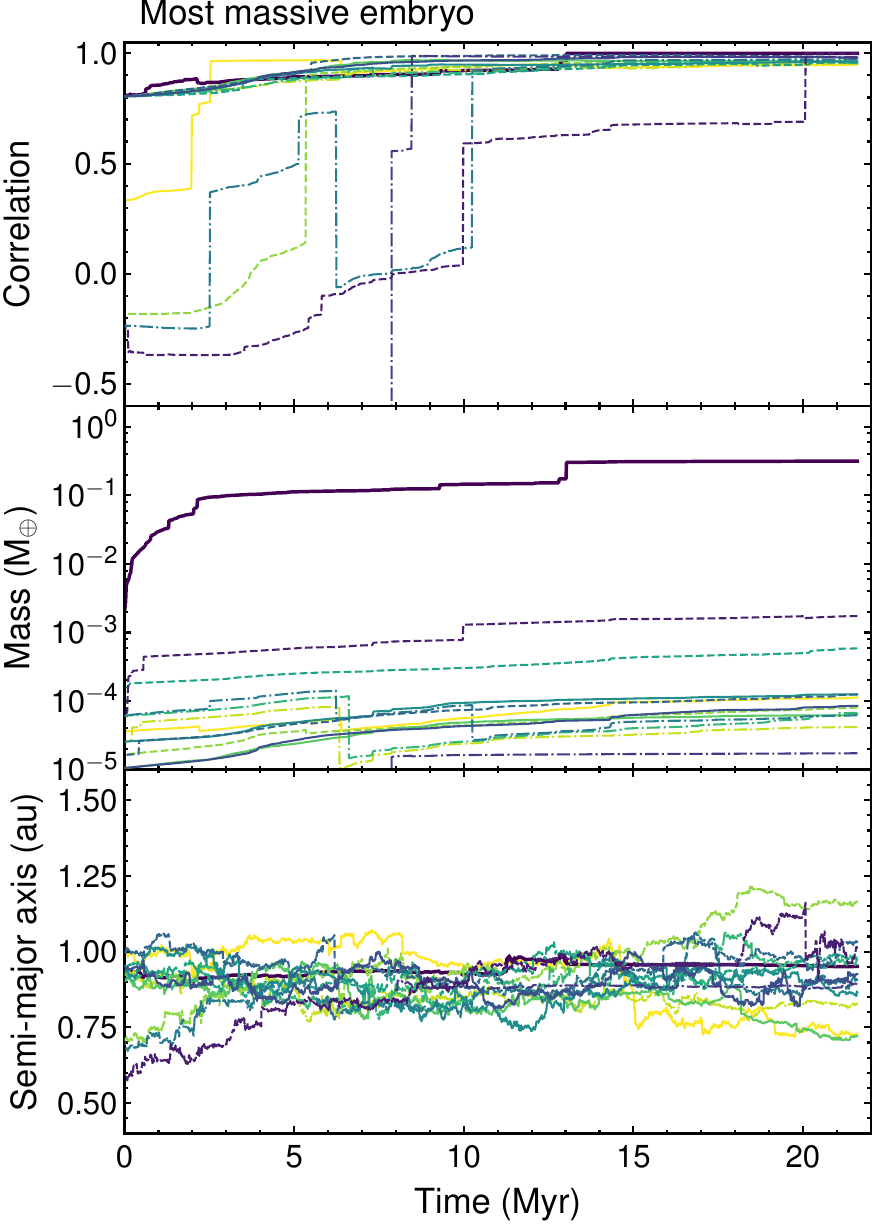}\hspace{8mm}
\centering\includegraphics[width=\columnwidth]{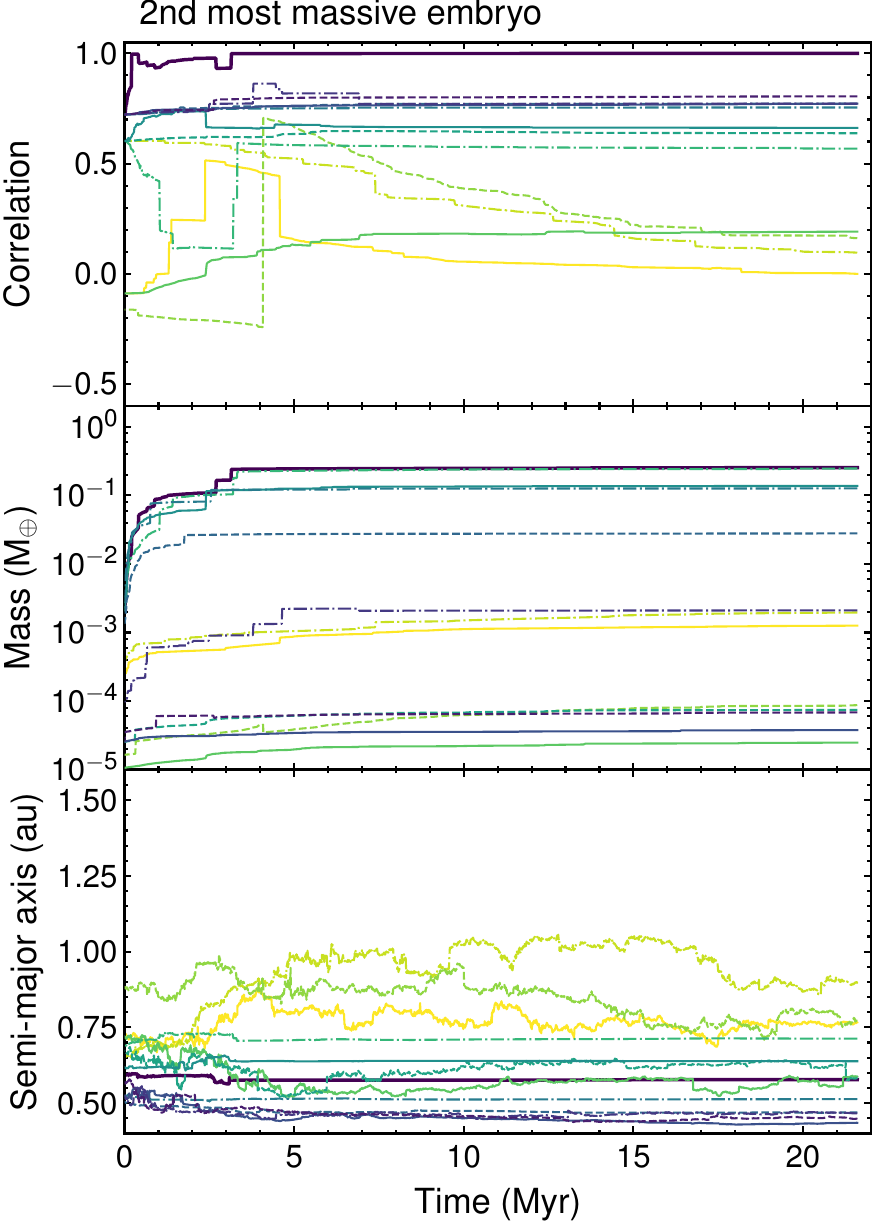}
\caption{Evolution of {provenance} correlation coefficient,  body mass, {and semi-major axis} for the growing embryo and the 12 bodies with the most similar {provenance}s at the end of an example calm accretion simulation (022f6sgas\_4). The {left} and {right} plots correspond to the most massive and second most massive embryos as shown in Figure \ref{f:embcomparecalm}. The thick dark purple line indicates the embryo and the thin solid, dashed, and dot-dash lines the most similar bodies, colored in order of correlation coefficient (purple -- highest -- to yellow -- lowest). \label{f:correlationevolutioncalm}}
\end{figure*}

The time evolution of body {provenance}s (Figures \ref{f:correlationevolutionGT} and \ref{f:correlationevolutioncalm}) allows us to examine when bodies become similar in {provenance} to each other. The 12 most-similar final bodies for the two largest embryos from the example Grand Tack simulation (shown in Figure \ref{f:embcompareGT}) are shown in Figure \ref{f:correlationevolutionGT}. The thick dark purple line indicates the growing embryo itself. By definition, an embryo has its ending {provenance} {at the end of the simulation run}, therefore its final correlation coefficient is 1.0. In both examples in Figure \ref{f:correlationevolutionGT}, the body that grew into the embryo (thick purple line) began with a correlation coefficient of 0.5 or less, indicating that the {provenance} of the embryo changed substantially as it grew. The most massive embryo ({left hand} panels) had several collisions with other embryos between 5 and 20\,Myr that significantly affected its {provenance}. The second most massive embryo ({right hand} panels) gained much of its {provenance} characteristics at $\sim$2\,Myr, during or shortly after Jupiter's migration event.

The majority of the most-similar planetesimals for the largest embryos in the Grand Tack (Figure \ref{f:correlationevolutionGT}) acquire their very-similar {provenance}s at the time they form (appearing as vertical lines beginning on the $x$-axis in {the upper} panels); {these similar planetesimals} are fragments ejected from the near-fully-grown planet. These fragments of proto-planets have typical masses ranging from $\sim$5\% to a few times the mass of the modern asteroid belt {($5\times10^{-4}$ \mearth, \citealt{DeMeo13})}. In the examples shown here, these surviving planetesimals can remain in the system for 10\,Myr or more as the planets continue growing. The most-similar bodies to the second largest embryo, shown in the {right hand} panels of Figure \ref{f:correlationevolutionGT}, are closer in {provenance} (higher correlation) at the end of the simulation run {than those for the most massive embryo.} {This greater similarity arises because} the second largest embryo gains its {(near-)}final {provenance} {and near-final mass} earlier than the most massive embryo ({left hand} panels), {which undergoes a greater change in provenance after the production of many of its most-similar planetesimals}.

{Some planetesimals exhibit substantial changes in provenance with small changes in mass, for example the purple dashed line at $\sim$20\,Myr in the left hand panels of Figure \ref{f:correlationevolutioncalm}. These sudden changes in provenance for an existing planetesimal are caused by collision with a more massive body. The remnants of the collision gain the mass-weighted average provenance of the impactors, and so a large difference in mass can cause a large change in the provenance of the smaller body.}

Embryos and planetesimals become  similar early {in the Grand Tack simulations} (in the first few Myr), as a result of collisions caused by the inward migration of Jupiter. There are thus many intermediate bodies (planetesimals and embryos) with similar {provenance}s close to the tack point early in the giant impact phase (see Figure \ref{f:embcompareGT}), and before the Moon-forming impact. One of these embryos could represent a Theia with an isotopic composition similar to the proto-Earth. The final planets can have similar {provenance}s to several intermediate embryos, and planetesimals could be ejected from any of these intermediate embryos.

There are several major differences in {provenance} evolution between the Grand Tack and calm accretion scenarios. Figure \ref{f:correlationevolutioncalm} shows the evolution of the two most massive embryos and the bodies most similar to them for an example calm accretion simulation. The embryos (thick dark purple lines) begin with {provenance}s much more similar to their final {provenance}s (correlation coefficients of $\sim$0.7 and 0.8) than seen for the Grand Tack example (Figure \ref{f:correlationevolutionGT}). With much less mixing than in the Grand Tack scenario, calm accretion results in embryos that are dominated by material that originated very close to their final locations.

\begin{figure}
\centering\includegraphics[width=\columnwidth]{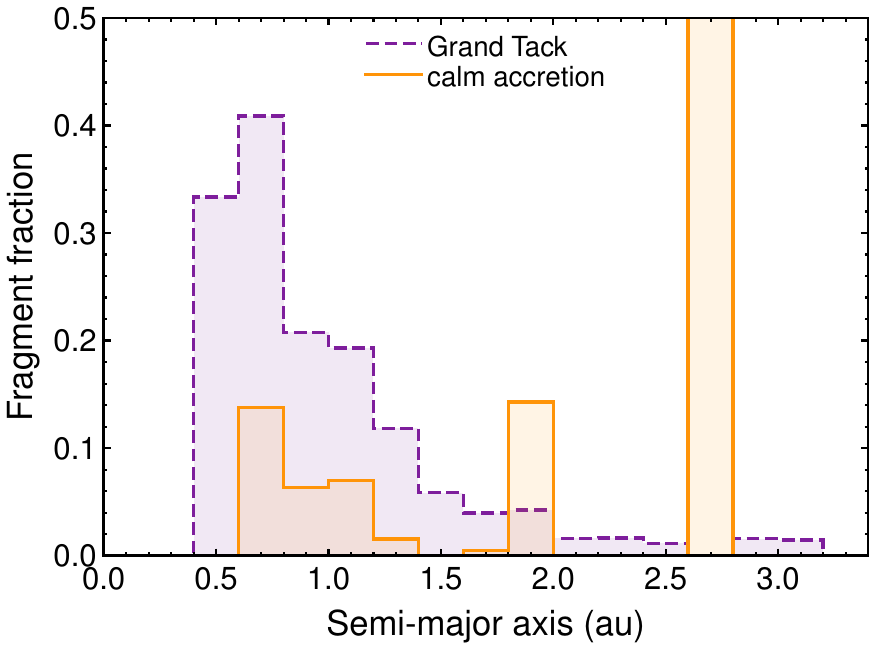}
\caption{{Fraction of surviving bodies that are collision fragments at the end of example Grand Tack (022GTJf6hgas, purple -- dashed line) and calm accretion (022f6sgas\_4, orange -- solid line) simulations. The large fragment fractions at 1.8--2.0 and 2.6--2.8\,au (0.14 and 1.0) for the calm accretion case are due to the small numbers of bodies in these bins (seven and one respectively). }\label{f:fragfrac}}
\end{figure}
The second clear difference in evolution of similar bodies in the calm accretion scenario is the origin of the most-similar smaller bodies. From Figure \ref{f:correlationevolutioncalm}, it is evident that most of these smaller bodies exist from the start of the simulation. Many of {the similar planetesimals} evolve in {provenance} very similarly to their similar embryo. Note, however, that many of the `most-similar' bodies for the second largest embryo have low correlations to the embryo's final {provenance}. Of the 24 planetesimals most-similar to the two largest embryos in the example shown in Figure \ref{f:correlationevolutioncalm}, only one is a clear collision fragment: the dot-dashed purple line appearing just before 8\,Myr in the upper {left} panels. Unlike the similar{-provenance} fragments in the Grand Tack case, this fragment was not ejected from the embryo; it appeared with a correlation coefficient of $\sim$0.5, and only gained its greater similarity later. {The calm accretion scenario results in fewer fragments than the Grand Tack simulations, and a smaller fraction of surviving bodies are fragments across the disk at the end of these simulations (Figure \ref{f:fragfrac}).}

\subsection{ Core provenance }

\begin{figure}
\centering\includegraphics[width=\columnwidth]{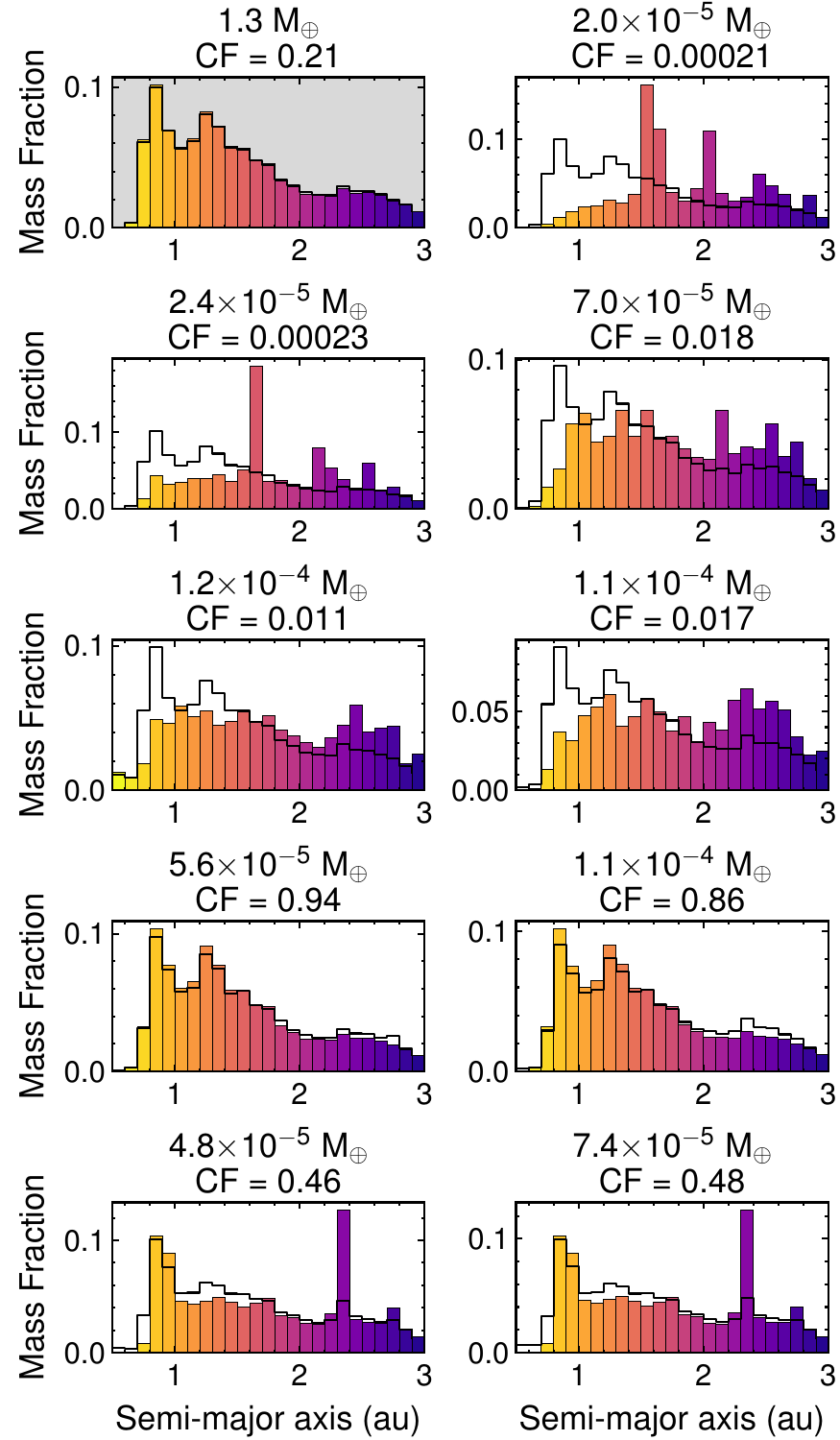}
\caption{{Provenance} histograms for the most massive embryo (grey background) and the nine bodies with the most similar bulk {provenance}s (with similarity decreasing left-to-right and top-to-bottom) at the end of a Grand Tack simulation (022GTJf6hgas). The black outline histogram shows the bulk {provenance}  for each body and the colored bars show the corresponding core {provenance}. The numbers above each panel give the mass and core mass fraction (CF) of each body.\label{f:corecomp}}
\end{figure}

We assume that all bodies in these simulations are differentiated and remain differentiated throughout. Since these bodies can undergo imperfect collisions, the {provenance}s of core and mantle can diverge. While this {decoupling} has a negligible effect on massive embryos, the planetesimals can develop substantial differences, particularly in the dynamically excited Grand Tack scenario  (see Figure \ref{f:corecomp}).

Approximately 9\% of bodies at the end of the Grand Tack simulations have correlation coefficients between their core and mantle {provenance}s below 0.5. 
The calm accretion simulations also result in planetesimals with dissimilar core and mantle, but these {planetesimals} represent a smaller fraction of the total number of bodies, $\sim$3\%.

For the planetesimals that have bulk {provenance}s very similar to an embryo but have low core mass fractions the metallic core can have a substantially different {provenance}  (see the upper rows in Figure \ref{f:corecomp}). In some cases the leftover planetesimals can have metallic cores similar to an embryo's {provenance}, while the silicate mantle (or bulk {provenance}) does not match.

This variation in core mass fraction and core and mantle {provenance}s means that small bodies can have non-chondritic metal-silicate ratios, and siderophile elements from different isotopic reservoirs than the lithophiles.


\section{Discussion}\label{s:discussion}

{\subsection{Disk evolution}}

{Jupiter's Grand Tack emplaces a large mass of planetary bodies from beyond $\sim$1.5\,au into the inner regions of the disk between $\sim$0.8 and $\sim$1.3\,au. The dynamical excitation and the increase in mass in the Earth-forming region lead to much faster evolution in the Grand Tack scenarios compared with the calm accretion scenarios. We can see from Figure \ref{f:massdist} that embryos have grown more massive more rapidly in the Grand Tack simulation. This difference in evolutionary timescales complicates the comparison between the two types of simulation.}

{The patterns of similar embryos and planetesimals seen in Figure \ref{f:nsimilar} are established early (in the first few Myr) and, in general, remain comparable throughout the rest of the simulation runs. The point in the disk at which the planetesimal distribution in the calm scenarios begins to rise toward the large numbers of similar planetesimals in the outer disk moves outward with time as the inside-out evolution proceeds. The similarity of planetesimals and embryos close to 1\,au shows significant variability with time in the Grand Tack scenario, likely due to the late production of collision fragments. If the calm accretion simulations were continued into the early stages of the giant impact phase we may see more similar-provenance fragments produced, though we expect the contrasting features of provenance similarity for the Grand Tack scenario to remain as these arise due to the dynamical excitation. }

{Our} simulations ignored planetesimals from the outer solar system (and the outer part of the inner solar system in the calm accretion case). While little outer solar system material is expected to be delivered to the inner disk in the calm accretion scenario, the migration of the giant planets in the Grand Tack model would have injected some outer solar system bodies into the Earth-formation region \citep[e.g.][]{Walsh11,Raymond17,CarterStewart20}. We expect that this outer solar system material would have been mixed into the inner disk and the growing embryos similarly to the purple and blue material from the outer parts of the inner disk. The leftover collisionally processed planetesimals then likely incorporated some of this outer solar system material similarly to the proto-Earth. The fates of outer solar system planetesimals, and the delivery of these, likely more volatile-rich, bodies to the Earth-forming region of the disk should be explored in future work.

{The simulations analyzed in this work (see Table \ref{t:sims}) cover different numerical resolutions (numbers of initial bodies) and nebular gas disk properties and evolution. We do not see substantial differences between the results for embryos in simulations with different resolutions, but there are much lower numbers of remaining planetesimal-sized bodies in the low resolution simulations (thus the low resolution simulations were not included in the lower panels of Figure \ref{f:nsimilar}) We also find that there are lower fractions of similar planetesimals for the low resolution simulations. The low resolution simulations have a higher mass limit, and thus, lack the smaller (less massive) resolved planetesimals. The planetesimal distributions in the low resolution simulations are thus more dominated by the larger planetesimals which are less likely to be fragments and more likely to survive provenance-altering collisions. We thus might expect to see greater numbers and fractions of similar planetesimals if the numerical resolution were further increased.} 

{One set of the calm accretion simulations used the same gas disk dissipation timescale as the Grand Tack simulations; the other set kept the MMSN gas constant throughout the simulation run. The simulations with nebular gas throughout have more similar embryos in the innermost regions of the disk principally because they produce greater numbers of small embryos on very cold orbits. There are fewer similar planetesimals at the end of these constant nebular gas simulations because they result in more efficient accretion of planetesimals, and some of those planetesimals that do survive in the inner regions of the disk have been transported from greater heliocentric distances due to the ongoing aerodynamic drag force. Overall, however, the differences between the two types of calm accretion simulations are small compared to the differences between calm and dynamically excited (Grand Tack) scenarios.}

{\subsection{Numerical limitations}}

As the simulations discussed in this work do not cover the entire giant impact phase, nor the final accretion of the leftover bodies, we cannot say which bodies survive until after the final giant impact on the Earth, or which are subsequently accreted. A large body of prior work on the giant impact stage and late accretion suggests that some significant portion of planetesimals survive, and that leftover small bodies are an important source of late-accreted mass \citep[e.g.][]{Bottke10,Schlichting12,Fischer18,Brasser20}. {However,} these models did not include the diversity of planetesimal {provenance}s we see. Full numerical simulations of planet growth, including both runaway growth and giant impacts, are still prohibitively computationally expensive, but will be required to fully develop our understanding of terrestrial planet accretion.

The simulations presented in this work ignored the generation of new planetesimals during the growth of planets. This simplification is applied near-universally in planet formation simulations. Improvements in our understanding of planetesimal formation, including recycling of collisionally-generated fragments, is needed in order to understand the effects on the compositional evolution of terrestrial planets.

To ensure the simulations were computationally feasible we treated metallic cores and silicate mantles as separate, non-interacting components: there was no chemical equilibration between cores and mantles during accretion. {Neglecting core-mantle equilibration} is a substantial simplification, particularly for smaller bodies that are less likely to have remained molten for long periods of time {and thus less likely to have experienced rapid, direct merging of accreted metals with their cores}. We expect differentiation and (re-)equilibration in a real system would increase the compositional similarity between mantle and core for moderately siderophile/lithophile elements. This simplification is, however, of little concern when considering HSEs, which partition strongly into the metal phase. Even with substantial metal-silicate equilibration, HSEs are expected to reside in the cores of accreted bodies and act as if they were a non-interacting component.

{\subsection{Moon formation}}

{Planetary bodies in our solar system have a wide range of measured isotopic compositions \citep[e.g.][]{Dauphas04,Scott18}, yet the Moon has an isotopic composition that is almost indistinguishable from that of the Earth \citep[e.g.][]{Wiechert01,Dauphas14}. In the canonical model for Moon formation with a Mars-mass impactor, most of the Moon is derived from the impactor, named Theia. Since such similarity would not be expected for two embryos that formed in different locations, it has been suggested that the Earth and Theia must have undergone extensive mixing during (or immediately after) the impact that led to the formation of the Moon \citep[e.g.][]{Pahlevan07,Canup12,Lock18}. }

We have  seen that in dynamically excited planet formation scenarios, there can be several smaller planetary embryos near the proto-Earth with {provenance}s similar to the proto-Earth during the early part of the giant impact phase (Figures \ref{f:embcompareGT} and \ref{f:nsimilar}). These {similar-provenance} embryos are potential Theias that would naturally have a similar stable isotopic composition to the proto-Earth before the Moon-forming impact. Even with significant variation in chemical and isotopic properties across the inner solar system, such a Theia would likely be sufficiently similar to alleviate some of the problems with the {isotopic} similarity of the Earth and Moon \citep[e.g.][]{Dauphas14}.

{In our current simulations, the potential Earth-like Theias are typically low-mass embryos (M $<$ 0.1\,\mearth) and, unless they grew substantially before impact with the proto-Earth, would likely require a high-velocity, high angular momentum impact in order to produce our Moon \citep[e.g][]{Cuk12,Lock18}. The more massive embryos (M $\gtrsim$ 0.1\,\mearth) expected for the canonical Moon-forming impact do not show a high similarity in provenance to our proto-Earths and would require extensive post-impact mixing. The small, similar-provenance embryos near the proto-Earth may, however, be advantageous for alternative giant impacts, including multiple events \citep[e.g.][]{Rufu17} and hit-and-run events \citep[e.g.][]{Asphaug21}. Similarity of source material in multiple planetary embryos would reduce the difficulties associated with explaining the Earth-Moon isotopic similarity.}

{\citet{Kaib15} studied oxygen isotope differences between Earth and Theia obtained from their $N$-body simulations with many possible isotopic gradients and distributions. Based on the measured difference between Earth and Mars, they found that it was very unlikely for Earth and Theia to have a sufficiently similar oxygen isotope composition to explain the observed similarity of the Earth and Moon. We did not find suitable Mars analogs in our smaller set of simulations, and thus cannot make a direct comparison to Mars. Further work is needed to determine whether dynamically excited planet formation can produce the final observed isotopic variations in the inner solar system.}

The similarity in stable isotopes (e.g.\ for oxygen) between the Earth and Moon would be expected if Theia {had a similar provenance} to the proto-Earth. {However, the stable isotope similarity with Earth is just one of several outstanding issues in lunar origin, including the large mass of the Moon, the similarity in Tungsten isotopes, the pattern of moderately volatile element depletion, and the inclination of the lunar orbit \citep{Lock20,Canup22}. Thus, factors other than material provenance must also be considered when discriminating between giant impact scenarios.}

{\subsection{Late accretion}}

{The HSE abundances in Earth's mantle suggest that a small percentage of the Earth's mass ($\sim$0.5\%) was delivered after the final giant impact \citep[e.g.][]{Walker09}. However, there is ongoing debate about the origin and composition of this late-accreted material \citep[e.g.][]{Dauphas17,Bermingham18,Hopp20}.}

We have seen that collisional evolution during the intermediate stages of terrestrial planet formation results in a range of planetesimal-sized bodies and small embryos with  similar {provenances} to the largest embryos. Some of these planetesimals will be accreted or ejected from the terrestrial planet region before the final giant impact, but it is expected that some will survive and remain in the inner disk \citep[e.g.][]{OBrien06,Schlichting12}.

Most of these `leftover' planetesimals are then expected to eventually be accreted by the terrestrial planets over the following billion years of evolution, or ejected from the solar system via dynamical interactions. A small fraction are likely scattered into the asteroid belt (Figure \ref{f:embcompareGT}), as has been suggested for iron-rich bodies \citep{Bottke06}, possibly becoming enstatite-like achondrites. However, we expect these `planet fragments' to represent a small fraction of the total mass in the asteroid belt (see Figure \ref{f:fragfrac}). Future work could use statistical evolution simulations to explore the delivery of these bodies to the asteroid belt.

Our results suggest that a portion of late-accreted mass came from fragments of intermediate planetary embryos ejected earlier in the history of the solar system, rather than primitive chondritic planetesimals. These leftover bodies naturally have similar {provenance}s to the planets. This leads us to a possible solution to the confounding origin of late-accreted mass: it is leftover material that already matches the Earth in composition.

As well as planetesimal-sized bodies with similar {provenance}s, and a small number that are almost compositionally identical to the proto-Earth, there are also many dissimilar planetesimals left at the end of the simulations (Figure \ref{f:embcompareGT}). We expect some mixture of these planetesimal-sized bodies to be accreted by the proto-Earth. Typically the more-similar bodies are close in the disk to the proto-Earth, the less-similar bodies are more distant {(see Figure \ref{f:hists})}, {and collision fragments are concentrated in the terrestrial planet formation region (see Figure \ref{f:fragfrac}).} The dissimilar planetesimals are also, typically, in the outer, less evolved regions of the inner solar system and thus, likely to decrease in number as evolution of the disk continues. We therefore consider the more-similar planetesimals more likely to end up accreted by the proto-Earth than the less-similar bodies.

{In this work, two bodies have similar provenances if they acquired the majority of their mass from the same regions of the disk. We assume that two bodies with identical provenance would have the same chemical and isotopic composition, at least for stable isotopes and strongly lithophile or siderophile elements. If the early protoplanetary disk was characterized by compositional gradients, two bodies that have different mixtures of material from different regions could end up with similar average isotopic compositions. For example, a mixture of yellow (0.6\,au) and blue (2.9\,au) material from our simulations would have an average color of magenta, corresponding to 1.75\,au, but such a mixed body would have a low provenance correlation with a body purely comprised of material originating at 1.75\,au. Bodies with low correlation may therefore still represent bodies with isotopically similar compositions, and there may be greater numbers of bodies that are a close compositional match to the proto-Earth than our results suggest. We note that current work on Earth's isotopic composition suggests that Earth is a mixture from different regions of the disk, rather than sourcing material from a single location \citep[e.g.][]{Fischer-Godde20,Burkhardt21}}

\begin{figure*}
\centering\includegraphics[width=\textwidth]{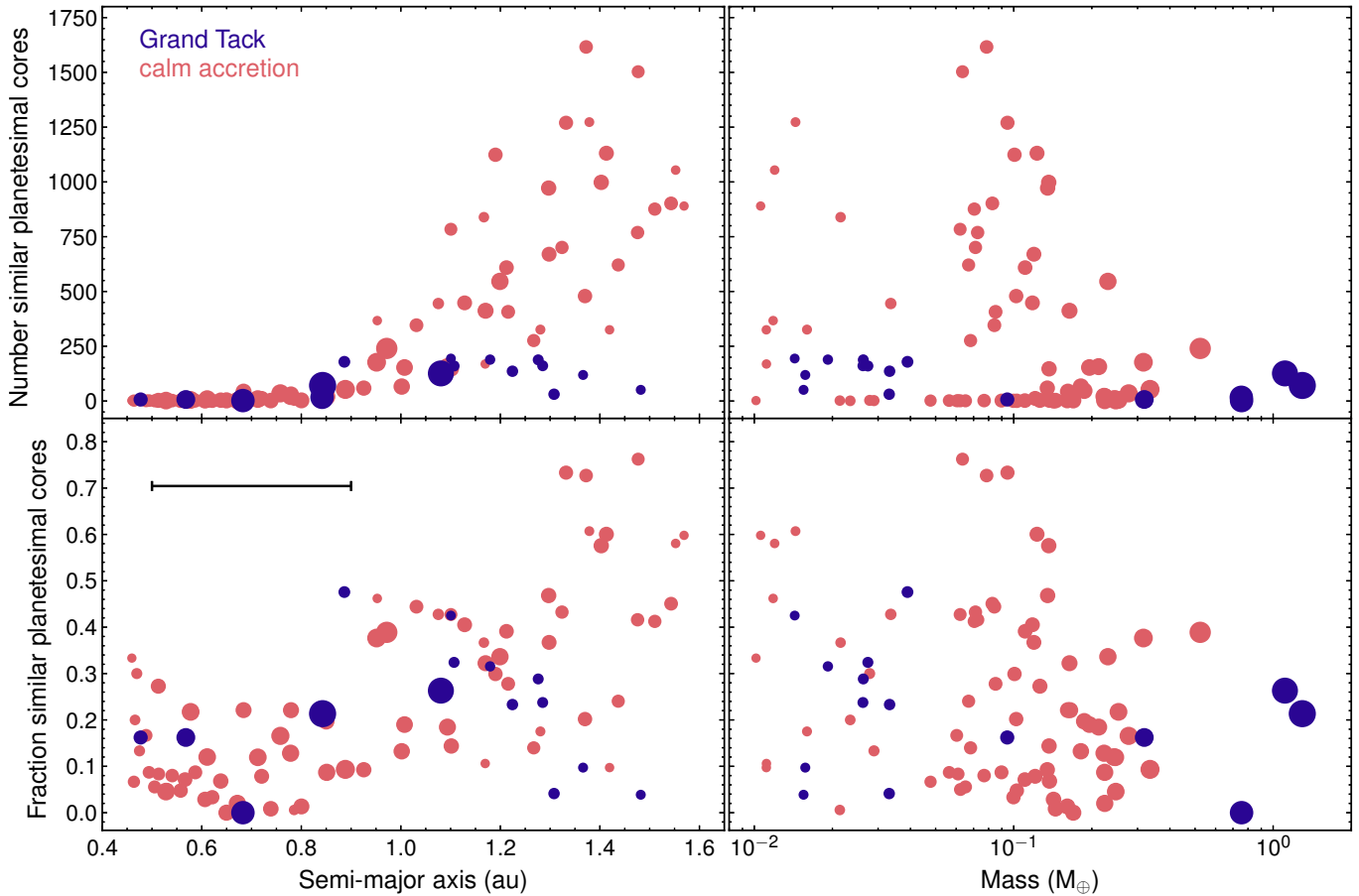}
\caption{{Number and fraction of planetesimals (M$<$0.01\,\mearth) with correlation coefficients between planetesimal core and embryo mantle of 0.5 or above for each embryo in several simulations. The numbers of planetesimals with similar cores are shown as a function of semi-major axis (left) and mass (right). Blue points are for Grand Tack simulations, magenta points are for calm accretion simulations. The sizes of the points are scaled to the mass of the embryos. Embryos from low resolution simulations are excluded. The fraction of planetesimals that are similar is calculated only for planetesimals with semi-major axes within 0.2\,au of the embryo, as indicated by the black horizontal bar in the lower left panel.} \label{f:nsimilarcore}}
\end{figure*}
The HSE Ruthenium has been the subject of several recent studies \citep[e.g.][]{Dauphas17,Bermingham18,Fischer-Godde20}. As Ru is highly siderophile it is expected that the majority of Ru in the accessible Earth was delivered via the late veneer. The Earth and the major chondrite groups appear to show a trend of increasingly s-process-depleted Ru (more negative $\varepsilon\,^{100}$Ru) with increasing heliocentric distance inferred for their formation \citep[e.g.][]{Fischer-Godde17}. The closest match to the Earth's Ru amongst known meteorites are the enstatite chondrites and IAB iron meteorites \citep{Bermingham18}. These Ru isotope {compositions} suggest that the late veneer was largely derived from inner solar system material with a similar isotopic composition to the enstatite chondrites and the proto-Earth \citep[e.g.][]{Dauphas17,Fischer-Godde17}. 

{Molybdenum, on the other hand, is moderately siderophile; while the early-delivered Mo is expected to have partitioned into the growing core, Mo delivered during the later stages of accretion would leave a signature in the mantle. Mo isotopes, therefore, trace the accretion of the final $\sim$10\% of Earth's mass \citep{Dauphas17,Burkhardt21}. Meteorites reveal that the isotopic compositions of Mo and Ru are correlated \citep[e.g.][]{Bermingham18}. Since the silicate Earth also lies on this `cosmic correlation,' both the final $\sim$10\% of Earth's mass and the late veneer must have originated from the same isotopic reservoir \citep[e.g.][]{Dauphas04,Bermingham18}, with no substantial change in the composition of impactors after core formation ceased.} 

Planetesimals or fragments from our simulations with {provenance}s similar to the proto-Earth could naturally explain a late veneer with {a matching} inner solar system isotopic signature. The compositional similarity between such late veneer impactors and the IAB and enstatite meteorites {(chondrites and achondrites)} is sensitive to the relative timing, and order, of the formation of the parent bodies of these meteorites and the migration of the giant planets.

{Figure \ref{f:nsimilar} shows that there are many planetesimals at the end of our simulations whose bulk provenances are similar to the final embryos in the Earth-forming region. When considering HSEs it is, however, the cores of potential late-accreted planetesimals that are most important. Figure \ref{f:nsimilarcore} shows the equivalent numbers and fractions of similar-provenance planetesimals using the correlation between embryo mantle and planetesimal core provenances. We see consistent patterns, more similar planetesimals in the Earth-forming annulus in the Grand Tack scenarios and an increasing number of similar planetesimals with increasing semi-major axis for the calm scenarios, and comparable, though in general slightly lower, numbers of similar planetesimals. Figure \ref{f:nsimilarcore} reveals a population of leftover planetesimals whose cores could supply Ru, and other HSEs, with inner solar system isotopic compositions matching the Earth's mantle.}

\citet{Fischer18Ru} explored Ruthenium and Molybdenum anomalies acquired by the terrestrial planets in $N$-body simulations using the perfect merging assumption. Differences in the duration of the simulations and the extent of the disk modeled preclude a direct comparison, but we can make some useful observations. In general terms we see similar trends, greater mixing and wider feeding zones in Grand Tack scenarios, and the requirement that the transition to `outer solar system composition Ru' occurred at large distances from the Sun, such that the Earth acquires {the correct Ru composition}. It is possible that the  similar{-provenance} collision fragments that we have found in our `imperfect accretion' simulations help to reduce this requirement for distant Ru-composition-transition-location by providing a further source for late accretion with {Earth-like Ru composition ($\varepsilon\,^{100}$Ru\,$\simeq$\,0)}. 

Recently, \citet{Fischer-Godde20} reported s-process-enriched Ru (with a positive $\varepsilon\,^{100}$Ru) present in rocks derived from early Archaean mantle ($\sim$3.8\,Gyr ago). This Ru isotopic signature reflects either preserved pre-late-veneer mantle or the composition of the early portion of late accretion. Mixing with an outer solar system, carbonaceous-chondrite-like component of the late veneer is favored in order to produce the Ru composition of the modern mantle \citep{Fischer-Godde20}. The {provenance}s resulting from our simulations could be compatible with these new Ru data if the leftover planetesimals represent the early component of late accretion, a carbonaceous-chondrite-like component were delivered later, and the parent bodies of enstatite chondrites formed from material with a sufficiently larger mean heliocentric distance than the Earth (`bluer' material in Figures \ref{f:massdist} and \ref{f:correlationevolutionGT}) thus giving them an intermediate isotopic composition.

The simulations we examined in this work show that planetesimal-sized bodies that may survive to the end of the giant impact phase have a range of core mass fractions, and that cores and mantles of these bodies can have different {provenance}s (radial origin distributions). This range of {properties} loosens the constraint on late veneer mass given by mantle HSEs. 
{The standard mass given for the late veneer (0.5\% of Earth's mass) is obtained via mass balance by matching the abundances of HSEs inferred for the Earth's mantle to a chondritic composition \citep[e.g.][]{Walker09}.}  
Bodies with metal-to-silicate ratios {higher} than chondritic would deliver {more} HSEs per unit mass than expected from bodies with chondritic compositions. Cores and mantles potentially having different {provenance}s complicates the interpretation of the isotopic signatures of late-accreted mass. Siderophile and lithophile elements delivered by any single body could possess isotopic signatures from different solar system reservoirs. Future simulations that consider only the giant impact phase of planet growth must consider the prior evolution of the small, planetesimal-sized bodies if the goal is to understand the compositions of the final planets.


\section{Summary}

Planet formation simulations result in a population of planetesimals left over from the early stages of accretion that {have similar provenances} to the proto-Earth. In dynamically excited scenarios, these bodies commonly originated as fragments of proto-planets ejected millions of years earlier. We find that dynamically excited planet formation models can result in a system of proto-planets and smaller embryos with very similar {provenance}s. These similarities in {provenance} suggest that Theia may have been isotopically similar to the proto-Earth if the protoplanetary disk was truncated and mixed via giant planet migration (or some other process).

Some of the planetesimal-sized bodies that exist during the early stages of the giant impact phase have large core mass fractions, with core {provenance}s very similar to that of the proto-Earth. If these planetesimals are later accreted by the planet, they would represent late-accreted mass with compositions that naturally match the composition of the Earth that can deliver HSEs to the mantle. The variation in core mass fraction and {provenance} amongst these leftover bodies complicates the constraints on late accretion provided by HSEs. Models of the giant impact phase of planet growth must consider the earlier evolution of the small bodies in order to understand the compositional evolution of the Earth.\\

\bigskip


The authors thank two anonymous reviewers for their careful reading and extremely helpful comments. PJC and STS acknowledge financial support from the Simons Foundation; PJC also acknowledges support from the Science and Technology Facilities council [grant number: ST/V000454/1]. 
This research has made use of NASA's Astrophysics Data System.\\ 

Data from simulations presented in this work are available from \citet{Carter22Data}. An executable version of this article, including scripts used to analyze simulation output and produce the figures found in this article, is available from \citet{Carter22Exec}. \\


\software{PKDGRAV \citep{Richardson00, Stadel01},  
          NumPy \citep{numpy}, 
          matplotlib \citep{matplotlib},
          Jupyter \citep{jupyter}
          }

\bibliography{references}{}

\begin{thebibliography}{}
\expandafter\ifx\csname natexlab\endcsname\relax\def\natexlab#1{#1}\fi
\providecommand{\url}[1]{\href{#1}{#1}}
\providecommand{\dodoi}[1]{doi:~\href{http://doi.org/#1}{\nolinkurl{#1}}}
\providecommand{\doeprint}[1]{\href{http://ascl.net/#1}{\nolinkurl{http://ascl.net/#1}}}
\providecommand{\doarXiv}[1]{\href{https://arxiv.org/abs/#1}{\nolinkurl{https://arxiv.org/abs/#1}}}

\bibitem[{{Adachi} {et~al.}(1976){Adachi}, {Hayashi}, \& {Nakazawa}}]{Adachi76}
{Adachi}, I., {Hayashi}, C., \& {Nakazawa}, K. 1976, Progress of Theoretical
  Physics, 56, 1756, \dodoi{10.1143/PTP.56.1756}

\bibitem[{{Asphaug} {et~al.}(2021){Asphaug}, {Emsenhuber}, {Cambioni},
  {Gabriel}, \& {Schwartz}}]{Asphaug21}
{Asphaug}, E., {Emsenhuber}, A., {Cambioni}, S., {Gabriel}, T. S.~J., \&
  {Schwartz}, S.~R. 2021, \psj, 2, 200, \dodoi{10.3847/PSJ/ac19b2}

\bibitem[{Bermingham {et~al.}(2018)Bermingham, Worsham, \&
  Walker}]{Bermingham18}
Bermingham, K., Worsham, E., \& Walker, R. 2018, Earth and Planetary Science
  Letters, 487, 221, \dodoi{https://doi.org/10.1016/j.epsl.2018.01.017}

\bibitem[{{Bonsor} {et~al.}(2015){Bonsor}, {Leinhardt}, {Carter}, {Elliott},
  {Walter}, \& {Stewart}}]{Bonsor15}
{Bonsor}, A., {Leinhardt}, Z.~M., {Carter}, P.~J., {et~al.} 2015, \icarus, 247,
  291, \dodoi{10.1016/j.icarus.2014.10.019}

\bibitem[{{Bottke} {et~al.}(2006){Bottke}, {Nesvorn{\'y}}, {Grimm},
  {Morbidelli}, \& {O'Brien}}]{Bottke06}
{Bottke}, W.~F., {Nesvorn{\'y}}, D., {Grimm}, R.~E., {Morbidelli}, A., \&
  {O'Brien}, D.~P. 2006, \nat, 439, 821, \dodoi{10.1038/nature04536}

\bibitem[{Bottke {et~al.}(2010)Bottke, Walker, Day, Nesvorny, \&
  Elkins-Tanton}]{Bottke10}
Bottke, W.~F., Walker, R.~J., Day, J. M.~D., Nesvorny, D., \& Elkins-Tanton, L.
  2010, Science, 330, 1527, \dodoi{10.1126/science.1196874}

\bibitem[{Brasser {et~al.}(2020)Brasser, Werner, \& Mojzsis}]{Brasser20}
Brasser, R., Werner, S., \& Mojzsis, S. 2020, Icarus, 338,
  \dodoi{10.1016/j.icarus.2019.113514}

\bibitem[{{Budde} {et~al.}(2016){Budde}, {Burkhardt}, {Brennecka},
  {Fischer-G{\"o}dde}, {Kruijer}, \& {Kleine}}]{Budde16}
{Budde}, G., {Burkhardt}, C., {Brennecka}, G.~A., {et~al.} 2016, Earth and
  Planetary Science Letters, 454, 293, \dodoi{10.1016/j.epsl.2016.09.020}

\bibitem[{Burkhardt {et~al.}(2011)Burkhardt, Kleine, Oberli, Pack, Bourdon, \&
  Wieler}]{Burkhardt11}
Burkhardt, C., Kleine, T., Oberli, F., {et~al.} 2011, Earth and Planetary
  Science Letters, 312, 390, \dodoi{https://doi.org/10.1016/j.epsl.2011.10.010}

\bibitem[{{Burkhardt} {et~al.}(2021){Burkhardt}, {Spitzer}, {Morbidelli},
  {Budde}, {Render}, {Kruijer}, \& {Kleine}}]{Burkhardt21}
{Burkhardt}, C., {Spitzer}, F., {Morbidelli}, A., {et~al.} 2021, Science
  Advances, 7, eabj7601, \dodoi{10.1126/sciadv.abj7601}

\bibitem[{{Cameron} \& {Ward}(1976)}]{Cameron+Ward}
{Cameron}, A.~G.~W., \& {Ward}, W.~R. 1976, in Lunar and Planetary Science
  Conference, Vol.~7, 120

\bibitem[{{Canup}(2012)}]{Canup12}
{Canup}, R.~M. 2012, Science, 338, 1052, \dodoi{10.1126/science.1226073}

\bibitem[{{Canup} {et~al.}(2021){Canup}, {Righter}, {Dauphas}, {Pahlevan},
  {{\'C}uk}, {Lock}, {Stewart}, {Salmon}, {Rufu}, {Nakajima}, \&
  {Magna}}]{Canup22}
{Canup}, R.~M., {Righter}, K., {Dauphas}, N., {et~al.} 2021, {New views on the
  Moon II}, arXiv:2103.02045.
\newblock \doarXiv{2103.02045}

\bibitem[{Carter(2022{\natexlab{a}})}]{Carter22Data}
Carter, P. 2022{\natexlab{a}}, {Replication Data for: ``Did Earth eat its
  leftovers? Impact ejecta as a component of the late veneer''},  Harvard
  Dataverse, \dodoi{10.7910/DVN/4NOILF}

\bibitem[{Carter(2022{\natexlab{b}})}]{Carter22Exec}
---. 2022{\natexlab{b}}, {PhilJCarter/EjectaCompositionLateAccretion: v1.0.0:
  Did Earth eat its leftovers? (accepted)}, v1.0.0,  Zenodo,
  \dodoi{10.5281/zenodo.6380403}

\bibitem[{{Carter} {et~al.}(2018){Carter}, {Leinhardt}, {Elliott}, {Stewart},
  \& {Walter}}]{Carter18}
{Carter}, P.~J., {Leinhardt}, Z.~M., {Elliott}, T., {Stewart}, S.~T., \&
  {Walter}, M.~J. 2018, Earth and Planetary Science Letters, 484, 276,
  \dodoi{10.1016/j.epsl.2017.12.012}

\bibitem[{{Carter} {et~al.}(2015){Carter}, {Leinhardt}, {Elliott}, {Walter}, \&
  {Stewart}}]{Carter15}
{Carter}, P.~J., {Leinhardt}, Z.~M., {Elliott}, T., {Walter}, M.~J., \&
  {Stewart}, S.~T. 2015, \apj, 813, 72, \dodoi{10.1088/0004-637X/813/1/72}

\bibitem[{Carter \& Stewart(2020)}]{CarterStewart20}
Carter, P.~J., \& Stewart, S.~T. 2020, The Planetary Science Journal, 1, 45,
  \dodoi{10.3847/psj/abaecc}

\bibitem[{{Chambers}(2001)}]{Chambers01}
{Chambers}, J.~E. 2001, \icarus, 152, 205, \dodoi{10.1006/icar.2001.6639}

\bibitem[{{Chambers}(2013)}]{Chambers13}
---. 2013, \icarus, 224, 43, \dodoi{10.1016/j.icarus.2013.02.015}

\bibitem[{{Chen} {et~al.}(2011){Chen}, {Lee}, {Lee}, {Jiun-San Shen}, \&
  {Chen}}]{Chen11}
{Chen}, H.-W., {Lee}, T., {Lee}, D.-C., {Jiun-San Shen}, J., \& {Chen}, J.-C.
  2011, \apjl, 743, L23, \dodoi{10.1088/2041-8205/743/1/L23}

\bibitem[{{Citron} \& {Stewart}(2022)}]{Citron22}
{Citron}, R.~I., \& {Stewart}, S.~T. 2022, The Planetary Science Journal, 3, in
  press. arXiv:2201.09349.
\newblock \doarXiv{2201.09349}

\bibitem[{{{\'C}uk} \& {Stewart}(2012)}]{Cuk12}
{{\'C}uk}, M., \& {Stewart}, S.~T. 2012, Science, 338, 1047,
  \dodoi{10.1126/science.1225542}

\bibitem[{{Dahl} \& {Stevenson}(2010)}]{Dahl10}
{Dahl}, T.~W., \& {Stevenson}, D.~J. 2010, Earth and Planetary Science Letters,
  295, 177, \dodoi{10.1016/j.epsl.2010.03.038}

\bibitem[{{Dauphas}(2017)}]{Dauphas17}
{Dauphas}, N. 2017, \nat, 541, 521, \dodoi{10.1038/nature20830}

\bibitem[{{Dauphas} {et~al.}(2014){Dauphas}, {Burkhardt}, {Warren}, \&
  {Teng}}]{Dauphas14}
{Dauphas}, N., {Burkhardt}, C., {Warren}, P., \& {Teng}, F.-Z. 2014,
  Philosophical Transactions of the Royal Society of London Series A, 372,
  2013.0244, \dodoi{10.1098/rsta.2013.0244}

\bibitem[{{Dauphas} {et~al.}(2004){Dauphas}, {Davis}, {Marty}, \&
  {Reisberg}}]{Dauphas04}
{Dauphas}, N., {Davis}, A.~M., {Marty}, B., \& {Reisberg}, L. 2004, Earth and
  Planetary Science Letters, 226, 465, \dodoi{10.1016/j.epsl.2004.07.026}

\bibitem[{Dauphas \& Morbidelli(2013)}]{Dauphas13}
Dauphas, N., \& Morbidelli, A. 2013, Treatise on Geochemistry. The Atmosphere:
  History, 6, \dodoi{10.1016/B978-0-08-095975-7.01301-2}

\bibitem[{{Dauphas} \& {Pourmand}(2011)}]{Dauphas11}
{Dauphas}, N., \& {Pourmand}, A. 2011, \nat, 473, 489,
  \dodoi{10.1038/nature10077}

\bibitem[{Davies {et~al.}(2020)Davies, Carter, Root, Kraus, Spaulding, Stewart,
  \& Jacobsen}]{Davies20}
Davies, E.~J., Carter, P.~J., Root, S., {et~al.} 2020, Journal of Geophysical
  Research: Planets, 125, \dodoi{10.1029/2019JE006227}

\bibitem[{{DeMeo} \& {Carry}(2013)}]{DeMeo13}
{DeMeo}, F.~E., \& {Carry}, B. 2013, \icarus, 226, 723,
  \dodoi{10.1016/j.icarus.2013.06.027}

\bibitem[{{Duncan} {et~al.}(1998){Duncan}, {Levison}, \& {Lee}}]{Duncan98}
{Duncan}, M.~J., {Levison}, H.~F., \& {Lee}, M.~H. 1998, \aj, 116, 2067,
  \dodoi{10.1086/300541}

\bibitem[{Fischer \& Nimmo(2018)}]{Fischer18}
Fischer, R.~A., \& Nimmo, F. 2018, Earth and Planetary Science Letters, 499,
  257, \dodoi{10.1016/j.epsl.2018.07.030}

\bibitem[{Fischer {et~al.}(2018)Fischer, Nimmo, \& O'Brien}]{Fischer18Ru}
Fischer, R.~A., Nimmo, F., \& O'Brien, D.~P. 2018, Earth and Planetary Science
  Letters, 482, 105, \dodoi{https://doi.org/10.1016/j.epsl.2017.10.055}

\bibitem[{{Fischer-G{\"o}dde} \& {Kleine}(2017)}]{Fischer-Godde17}
{Fischer-G{\"o}dde}, M., \& {Kleine}, T. 2017, \nat, 541, 525,
  \dodoi{10.1038/nature21045}

\bibitem[{Fischer-G{\"o}dde {et~al.}(2020)Fischer-G{\"o}dde, Elfers,
  M{\"u}nker, Szilas, Maier, Messling, Morishita, Van~Kranendonk, \&
  Smithies}]{Fischer-Godde20}
Fischer-G{\"o}dde, M., Elfers, B.-M., M{\"u}nker, C., {et~al.} 2020, Nature,
  579, 240, \dodoi{10.1038/s41586-020-2069-3}

\bibitem[{Genda {et~al.}(2017)Genda, Brasser, \& Mojzsis}]{Genda17}
Genda, H., Brasser, R., \& Mojzsis, S. 2017, Earth and Planetary Science
  Letters, 480, 25, \dodoi{10.1016/j.epsl.2017.09.041}

\bibitem[{{Genda} {et~al.}(2017){Genda}, {Iizuka}, {Sasaki}, {Ueno}, \&
  {Ikoma}}]{Genda17_GIFs}
{Genda}, H., {Iizuka}, T., {Sasaki}, T., {Ueno}, Y., \& {Ikoma}, M. 2017, Earth
  and Planetary Science Letters, 470, 87, \dodoi{10.1016/j.epsl.2017.04.035}

\bibitem[{{Greenwood} {et~al.}(2018){Greenwood}, {Barrat}, {Miller}, {Anand},
  {Dauphas}, {Franchi}, {Sillard}, \& {Starkey}}]{Greenwood18}
{Greenwood}, R.~C., {Barrat}, J.-A., {Miller}, M.~F., {et~al.} 2018, Science
  Advances, 4, eaao5928, \dodoi{10.1126/sciadv.aao5928}

\bibitem[{Halliday(2013)}]{Halliday13}
Halliday, A.~N. 2013, Geochimica et Cosmochimica Acta, 105, 146,
  \dodoi{10.1016/j.gca.2012.11.015}

\bibitem[{{Hartmann} \& {Davis}(1975)}]{Hartmann+Davis}
{Hartmann}, W.~K., \& {Davis}, D.~R. 1975, \icarus, 24, 504,
  \dodoi{10.1016/0019-1035(75)90070-6}

\bibitem[{{Hopp} {et~al.}(2020){Hopp}, {Budde}, \& {Kleine}}]{Hopp20}
{Hopp}, T., {Budde}, G., \& {Kleine}, T. 2020, Earth and Planetary Science
  Letters, 534, 116065, \dodoi{10.1016/j.epsl.2020.116065}

\bibitem[{{Hunter}(2007)}]{matplotlib}
{Hunter}, J.~D. 2007, Computing in Science \& Engineering, 9, 90,
  \dodoi{10.1109/MCSE.2007.55}

\bibitem[{{Kaib} \& {Cowan}(2015)}]{Kaib15}
{Kaib}, N.~A., \& {Cowan}, N.~B. 2015, \icarus, 252, 161,
  \dodoi{10.1016/j.icarus.2015.01.013}

\bibitem[{Kendall \& Melosh(2016)}]{Kendall16}
Kendall, J.~D., \& Melosh, H. 2016, Earth and Planetary Science Letters, 448,
  24, \dodoi{https://doi.org/10.1016/j.epsl.2016.05.012}

\bibitem[{Kluyver {et~al.}(2016)Kluyver, Ragan-Kelley, P{\'e}rez, Granger,
  Bussonnier, Frederic, Kelley, Hamrick, Grout, Corlay, Ivanov, Avila, Abdalla,
  Willing, \& development team}]{jupyter}
Kluyver, T., Ragan-Kelley, B., P{\'e}rez, F., {et~al.} 2016, in Positioning and
  Power in Academic Publishing: Players, Agents and Agendas, ed. F.~Loizides \&
  B.~Scmidt (Netherlands: IOS Press), 87--90.
\newblock \url{https://eprints.soton.ac.uk/403913/}

\bibitem[{{Kokubo} \& {Ida}(1996)}]{Kokubo+Ida96}
{Kokubo}, E., \& {Ida}, S. 1996, \icarus, 123, 180,
  \dodoi{10.1006/icar.1996.0148}

\bibitem[{{Kokubo} \& {Ida}(1998)}]{Kokubo+Ida98}
---. 1998, \icarus, 131, 171, \dodoi{10.1006/icar.1997.5840}

\bibitem[{{Kokubo} \& {Ida}(2002)}]{Kokubo+Ida02}
---. 2002, \apj, 581, 666, \dodoi{10.1086/344105}

\bibitem[{{Kruijer} {et~al.}(2014){Kruijer}, {Touboul}, {Fischer-G{\"o}dde},
  {Bermingham}, {Walker}, \& {Kleine}}]{Kruijer14}
{Kruijer}, T.~S., {Touboul}, M., {Fischer-G{\"o}dde}, M., {et~al.} 2014,
  Science, 344, 1150, \dodoi{10.1126/science.1251766}

\bibitem[{{Leinhardt} {et~al.}(2015){Leinhardt}, {Dobinson}, {Carter}, \&
  {Lines}}]{Leinhardt15}
{Leinhardt}, Z.~M., {Dobinson}, J., {Carter}, P.~J., \& {Lines}, S. 2015, \apj,
  806, 23, \dodoi{10.1088/0004-637X/806/1/23}

\bibitem[{{Leinhardt} \& {Richardson}(2005)}]{Leinhardt05}
{Leinhardt}, Z.~M., \& {Richardson}, D.~C. 2005, \apj, 625, 427,
  \dodoi{10.1086/429402}

\bibitem[{{Leinhardt} \& {Stewart}(2012)}]{Leinhardt12}
{Leinhardt}, Z.~M., \& {Stewart}, S.~T. 2012, \apj, 745, 79,
  \dodoi{10.1088/0004-637X/745/1/79}

\bibitem[{{Lewis}(1974)}]{Lewis74}
{Lewis}, J.~S. 1974, Science, 186, 440, \dodoi{10.1126/science.186.4162.440}

\bibitem[{{Liu} \& {Ji}(2020)}]{Liu20}
{Liu}, B., \& {Ji}, J. 2020, Research in Astronomy and Astrophysics, 20, 164,
  \dodoi{10.1088/1674-4527/20/10/164}

\bibitem[{{Lock} {et~al.}(2020){Lock}, {Bermingham}, {Parai}, \&
  {Boyet}}]{Lock20}
{Lock}, S.~J., {Bermingham}, K.~R., {Parai}, R., \& {Boyet}, M. 2020, \ssr,
  216, 109, \dodoi{10.1007/s11214-020-00729-z}

\bibitem[{{Lock} {et~al.}(2018){Lock}, {Stewart}, {Petaev}, {Leinhardt},
  {Mace}, {Jacobsen}, \& {Cuk}}]{Lock18}
{Lock}, S.~J., {Stewart}, S.~T., {Petaev}, M.~I., {et~al.} 2018, Journal of
  Geophysical Research (Planets), 123, 910, \dodoi{10.1002/2017JE005333}

\bibitem[{{Mann} {et~al.}(2012){Mann}, {Frost}, {Rubie}, {Becker}, \&
  {Aud{\'e}tat}}]{Mann12}
{Mann}, U., {Frost}, D.~J., {Rubie}, D.~C., {Becker}, H., \& {Aud{\'e}tat}, A.
  2012, \gca, 84, 593, \dodoi{10.1016/j.gca.2012.01.026}

\bibitem[{{Marcus} {et~al.}(2010){Marcus}, {Sasselov}, {Stewart}, \&
  {Hernquist}}]{Marcus10}
{Marcus}, R.~A., {Sasselov}, D., {Stewart}, S.~T., \& {Hernquist}, L. 2010,
  \apjl, 719, L45, \dodoi{10.1088/2041-8205/719/1/L45}

\bibitem[{Marty {et~al.}(2017)Marty, Altwegg, Balsiger, Bar-Nun, Bekaert,
  Berthelier, Bieler, Briois, Calmonte, Combi, De~Keyser, Fiethe, Fuselier,
  Gasc, Gombosi, Hansen, H{\"a}ssig, J{\"a}ckel, Kopp, Korth, Le~Roy, Mall,
  Mousis, Owen, R{\`e}me, Rubin, S{\'e}mon, Tzou, Waite, \& Wurz}]{Marty17}
Marty, B., Altwegg, K., Balsiger, H., {et~al.} 2017, Science, 356, 1069,
  \dodoi{10.1126/science.aal3496}

\bibitem[{{Morbidelli} \& {Crida}(2007)}]{Morbi07}
{Morbidelli}, A., \& {Crida}, A. 2007, \icarus, 191, 158,
  \dodoi{10.1016/j.icarus.2007.04.001}

\bibitem[{{O'Brien} {et~al.}(2006){O'Brien}, {Morbidelli}, \&
  {Levison}}]{OBrien06}
{O'Brien}, D.~P., {Morbidelli}, A., \& {Levison}, H.~F. 2006, \icarus, 184, 39,
  \dodoi{10.1016/j.icarus.2006.04.005}

\bibitem[{{O'Neill}(1991)}]{ONeill91}
{O'Neill}, H. S.~C. 1991, \gca, 55, 1159, \dodoi{10.1016/0016-7037(91)90169-6}

\bibitem[{{Pahlevan} \& {Stevenson}(2007)}]{Pahlevan07}
{Pahlevan}, K., \& {Stevenson}, D.~J. 2007, Earth and Planetary Science
  Letters, 262, 438, \dodoi{10.1016/j.epsl.2007.07.055}

\bibitem[{Qin {et~al.}(2010)Qin, Alexander, Carlson, Horan, \&
  Yokoyama}]{Qin09}
Qin, L., Alexander, C.~M., Carlson, R.~W., Horan, M.~F., \& Yokoyama, T. 2010,
  Geochimica et Cosmochimica Acta, 74, 1122,
  \dodoi{https://doi.org/10.1016/j.gca.2009.11.005}

\bibitem[{{Raymond} \& {Izidoro}(2017)}]{Raymond17}
{Raymond}, S.~N., \& {Izidoro}, A. 2017, \icarus, 297, 134,
  \dodoi{10.1016/j.icarus.2017.06.030}

\bibitem[{{Raymond} {et~al.}(2009){Raymond}, {O'Brien}, {Morbidelli}, \&
  {Kaib}}]{Raymond09}
{Raymond}, S.~N., {O'Brien}, D.~P., {Morbidelli}, A., \& {Kaib}, N.~A. 2009,
  \icarus, 203, 644, \dodoi{10.1016/j.icarus.2009.05.016}

\bibitem[{Raymond {et~al.}(2013)Raymond, Schlichting, Hersant, \&
  Selsis}]{Raymond13}
Raymond, S.~N., Schlichting, H.~E., Hersant, F., \& Selsis, F. 2013, Icarus,
  226, 671, \dodoi{https://doi.org/10.1016/j.icarus.2013.06.019}

\bibitem[{{Reufer} {et~al.}(2012){Reufer}, {Meier}, {Benz}, \&
  {Wieler}}]{Reufer12}
{Reufer}, A., {Meier}, M. M.~M., {Benz}, W., \& {Wieler}, R. 2012, \icarus,
  221, 296, \dodoi{10.1016/j.icarus.2012.07.021}

\bibitem[{{Richardson} {et~al.}(2000){Richardson}, {Quinn}, {Stadel}, \&
  {Lake}}]{Richardson00}
{Richardson}, D.~C., {Quinn}, T., {Stadel}, J., \& {Lake}, G. 2000, \icarus,
  143, 45, \dodoi{10.1006/icar.1999.6243}

\bibitem[{{Rubie} {et~al.}(2016){Rubie}, {Laurenz}, {Jacobson}, {Morbidelli},
  {Palme}, {Vogel}, \& {Frost}}]{Rubie16}
{Rubie}, D.~C., {Laurenz}, V., {Jacobson}, S.~A., {et~al.} 2016, Science, 353,
  1141, \dodoi{10.1126/science.aaf6919}

\bibitem[{{Rufu} {et~al.}(2017){Rufu}, {Aharonson}, \& {Perets}}]{Rufu17}
{Rufu}, R., {Aharonson}, O., \& {Perets}, H.~B. 2017, Nature Geoscience, 10,
  89, \dodoi{10.1038/ngeo2866}

\bibitem[{Schlichting {et~al.}(2012)Schlichting, Warren, \&
  Yin}]{Schlichting12}
Schlichting, H.~E., Warren, P.~H., \& Yin, Q.-Z. 2012, The Astrophysical
  Journal, 752, 8, \dodoi{10.1088/0004-637x/752/1/8}

\bibitem[{{Scott} {et~al.}(2018){Scott}, {Krot}, \& {Sanders}}]{Scott18}
{Scott}, E.~R.~D., {Krot}, A.~N., \& {Sanders}, I.~S. 2018, \apj, 854, 164,
  \dodoi{10.3847/1538-4357/aaa5a5}

\bibitem[{{Stadel}(2001)}]{Stadel01}
{Stadel}, J.~G. 2001, PhD thesis, UNIVERSITY OF WASHINGTON

\bibitem[{{Stewart} \& {Leinhardt}(2012)}]{Stewart12}
{Stewart}, S.~T., \& {Leinhardt}, Z.~M. 2012, \apj, 751, 32,
  \dodoi{10.1088/0004-637X/751/1/32}

\bibitem[{{S{\"u}li}(2021)}]{Suli21}
{S{\"u}li}, {\'A}. 2021, \mnras, 503, 4700, \dodoi{10.1093/mnras/stab669}

\bibitem[{Touboul {et~al.}(2007)Touboul, Kleine, Bourdon, Palme, \&
  Wieler}]{Touboul07}
Touboul, M., Kleine, T., Bourdon, B., Palme, H., \& Wieler, R. 2007, Nature,
  450, 1206, \dodoi{10.1038/nature06428}

\bibitem[{{Trinquier} {et~al.}(2009){Trinquier}, {Elliott}, {Ulfbeck}, {Coath},
  {Krot}, \& {Bizzarro}}]{Trinquier09}
{Trinquier}, A., {Elliott}, T., {Ulfbeck}, D., {et~al.} 2009, Science, 324,
  374, \dodoi{10.1126/science.1168221}

\bibitem[{Van Der~Walt {et~al.}(2011)Van Der~Walt, Colbert, \&
  Varoquaux}]{numpy}
Van Der~Walt, S., Colbert, S.~C., \& Varoquaux, G. 2011, Computing in Science
  \& Engineering, 13, 22

\bibitem[{Walker(2009)}]{Walker09}
Walker, R.~J. 2009, Geochemistry, 69, 101,
  \dodoi{https://doi.org/10.1016/j.chemer.2008.10.001}

\bibitem[{{Walsh} \& {Levison}(2019)}]{Walsh19}
{Walsh}, K.~J., \& {Levison}, H.~F. 2019, \icarus, 329, 88,
  \dodoi{10.1016/j.icarus.2019.03.031}

\bibitem[{{Walsh} {et~al.}(2011){Walsh}, {Morbidelli}, {Raymond}, {O'Brien}, \&
  {Mandell}}]{Walsh11}
{Walsh}, K.~J., {Morbidelli}, A., {Raymond}, S.~N., {O'Brien}, D.~P., \&
  {Mandell}, A.~M. 2011, \nat, 475, 206, \dodoi{10.1038/nature10201}

\bibitem[{Warren(2011)}]{Warren11}
Warren, P.~H. 2011, Earth and Planetary Science Letters, 311, 93,
  \dodoi{https://doi.org/10.1016/j.epsl.2011.08.047}

\bibitem[{{Wiechert} {et~al.}(2001){Wiechert}, {Halliday}, {Lee}, {Snyder},
  {Taylor}, \& {Rumble}}]{Wiechert01}
{Wiechert}, U., {Halliday}, A.~N., {Lee}, D.~C., {et~al.} 2001, Science, 294,
  345, \dodoi{10.1126/science.1063037}

\bibitem[{{Wood} {et~al.}(2006){Wood}, {Walter}, \& {Wade}}]{Wood06}
{Wood}, B.~J., {Walter}, M.~J., \& {Wade}, J. 2006, \nat, 441, 825,
  \dodoi{10.1038/nature04763}

\end{thebibliography}
\bibliographystyle{aasjournal}

\end{document}